\documentclass[iop]{emulateapj}
\usepackage{paralist}
\shorttitle{Magnetized massive disks}
\shortauthors{M-K. Lin}

\usepackage{amsmath}
\usepackage{bm}
\newcommand{\p}{\partial}

\newcommand{\sbar}{\bar{\sigma}} 
\newcommand{\imgi}{\mathrm{i}}

\newcommand{\avg}[1]{\langle{#1}\rangle}

\newcommand{\dd}{\delta}
\newcommand{\real}{\operatorname{Re}}

\newcommand{\zmax}{Z_s}

\newcommand{\csmid}{c_{s0}}

\newcommand{\dvx}{\dd v_x}
\newcommand{\dvy}{\dd v_y}
\newcommand{\dvz}{\dd v_z}
\newcommand{\dbx}{\dd B_x}
\newcommand{\dby}{\dd B_y}
\newcommand{\dbz}{\dd B_z}
\newcommand{\w}{ \widetilde{W}}
\newcommand{\dphi}{\dd \Phi}

\defcitealias{lin12}{L12} 

\begin{document}

\title{Linear stability of magnetized massive protoplanetary disks}

\author{Min-Kai Lin }%(+Menou, Fromang, Youdin)}
\affil{Canadian Institute for Theoretical Astrophysics,
  60 St. George Street, Toronto, ON, M5S 3H8, Canada}
\email{mklin924@cita.utoronto.ca}

\begin{abstract}
  Magneto-rotational instability (MRI) and gravitational instability
  (GI) are the two principle routes to turbulent angular momentum
  transport in accretion disks. Protoplanetary disks may develop 
  both. This paper aims to reinvigorate interest in the study of
  magnetized  massive protoplanetary disks, starting from the basic
  issue of stability. The local linear stability of a 
  self-gravitating, uniformly magnetized, differentially rotating, 
  three-dimensional stratified disk subject to axisymmetric perturbations is
  calculated numerically. The formulation includes 
  resistivity. It is found that the reduction in the disk thickness
  by self-gravity can decrease MRI growth rates; the MRI
  becomes global in the vertical direction, and MRI modes 
  with small radial length scales are stabilized. The
    maximum vertical field strength that permits the MRI
    in a strongly self-gravitating polytropic disk  with polytropic index $\Gamma=1$ 
    is estimated to be $B_{z,\mathrm{max}} \simeq \csmid\Omega\sqrt{\mu_0/16\pi G} $, where $\csmid$ is the midplane sound speed and $\Omega$ is the angular velocity. 
  In massive
  disks with layered resistivity, the MRI is not well-localized
  to regions where the Elsasser number exceeds unity. For MRI modes
  with radial length scales on the order of the disk thickness, self-gravity 
  can enhance density perturbations, an effect that becomes 
  significant in the presence of a strong toroidal field, and
    which depends on the symmetry of the underlying MRI mode. In
  gravitationally unstable disks where GI and MRI growth rates are
  comparable, the character of unstable modes can transition smoothly
  between MRI and GI. Implications for non-linear simulations are
  discussed briefly.   

%layered dis   
  %Addition effects such as
  %resistivity and azimuthal fields are considered. 
%We examine the local linear stability of
%  magnetized, self-gravitating and differentially rotating disks to
%  axisymmetric perturbations. 
%
%not many studies model both
%revisits 
\end{abstract}

\section{Introduction}
Astrophysical disks host a wide range of fluid instabilities. Among
them, the magneto-rotational instability \citep[MRI,][]{chandrasekhar61,
  balbus91,balbus98} 
and gravitational instability \citep[GI, ][]{toomre64,goldreich65a,goldreich65b} 
provide robust pathways to turbulent angular momentum transport that
enables mass accretion \citep[][ 
  and references therein]{balbus99,armitage11,turner14}.   
They are also relevant to planet formation theory. For example,
the strength of MRI turbulence directly affect planetesimal dynamics
in protoplanetary disks \citep{yang12,gressel12}; while GI can
potentially form giant planets directly through disk fragmentation
\citep{boss97,boss98,gammie01,voro13,helled14}.

Accretion disks such as those surrounding black holes can develop both
MRI and GI \citep{menou01,goodman03}.  
Protoplanetary disks (PPDs) are also expected to be massive and  
magnetized in its earliest evolutionary phase \citep{inutsuka10}.  
The interplay between MRI and GI has been invoked to explain outbursts
in circumstellar disks \citep{armitage01, zhu10a, zhu10b,
  martin12b}, and predicts similar phenomenon in circumplanetary disks
\citep{lubow12}. This results from the development of `dead zones' 
--- magnetically inactive, laminar regions near the disk midplane --- with
magnetized layers above and below 
\citep{gammie96,martin12,landry13}. Mass accumulation in the dead zone can lead to
GI and trigger MRI through heating. In these models, the condition required
for MRI is realized through GI, but the MRI is unaffected by disk
self-gravity.

PPDs subject to both MRI and GI are often modeled through separate
turbulent viscosity coefficients in a hydrodynamical framework 
\citep{terquem08}. This implicitly assumes that the development of MRI 
and GI can be assessed independently. 
Circumstellar disk models that explicitly combine the equations of magneto-hydrodynamics
and self-gravity have been limited to a few early simulations 
\citep{fromang04,fromang04b,fromang05}. It will be necessary to 
revisit and extend these pioneering calculations to fully explore the
impact of MRI and GI on the structure and evolution of PPDs. In
preparation of this, it is important to have a thorough understanding
of the stability properties of such systems.

Since compressibility is not fundamental for the MRI, much of the
early stability calculations assume incompressible perturbations
\citep{goodman94,jin96}. However, recent works indicate
compressibility may be important under certain conditions,
such as strong fields \citep{kim00, pessah05,bonanno07}.   
Previous MRI studies have also focused on modes with vanishing
radial wavenumber, because they are the most unstable
\citep{sano99,reyes01}. Self-gravity has minimal effect on such
perturbations in a rotating disk. However, modes
with radial length scales on the order of the disk scale height may be
subject to self-gravity. It is therefore of interest to generalize the MRI with
non-zero radial wavenumbers to massive disks.

The effect of a magnetic field on the GI of rotating disks has been
considered recently by \cite{lizano10}, who generalized the Toomre stability
criterion for razor-thin disks to include a vertical field. For
circumstellar disks, the authors concluded that the field is
stabilizing. This is consistent with previous analysis by
\cite{nakamura83} for three-dimensional (3D) uniformly rotating
disks. However, the GI of 3D differentially rotating disks have
mostly neglected magnetic fields \citep{mamat10,kim12}, but such disks
are subject to the MRI if magnetized.

This work marks the beginning of our study of magnetized,
self-gravitating PPDs. We start from linear calculations, which  
have the advantage that a wide range of parameters can be studied at
negligible computational cost. This  allows us to identify
conditions, if any, under which MRI and GI 
cannot be considered independent. It is also important to have such 
calculations to benchmark and guide future non-linear
simulations.

This paper is organized as follows. \S\ref{setup} lists the governing
equations and describes the disk equilibria under consideration. 
The linear problem is formulated in \S\ref{linear}. The impact
of self-gravity on the MRI with a vertical field is discussed in
\S\ref{result1}, gravitationally unstable disks are considered in
\S\ref{result2}, and equilibria including an azimuthal field is explored in
\S\ref{result3}. We summarize results in \S\ref{summary} with a 
discussion of important extensions to our current models.

\section{Local disk model}\label{setup}
We study the local stability of an inviscid, self-gravitating and
magnetized fluid disk orbiting a central star with
potential $\Phi_*(r,z)$, where $(r,\varphi,z)$ are cylindrical
co-ordinates from the star. We use the shearing box approximation     
\citep{goldreich65b} to consider a small patch of the disk at
a fiducial radius $r=r_0$. The local frame rotates at angular velocity 
$\Omega_0=\Omega(r_0,0)$ about the star, where $r\Omega^2 =
\p\Phi_*/\p r$. We also define $S\equiv-r\p\Omega/\p r$ as the local shear
rate and $\Omega_z^2\equiv\p^2\Phi_*/\p z^2$ as the square of the
local vertical frequency. 

A Cartesian co-ordinate system $(x,y,z)$ is set
up in this local frame, corresponding to the radial, azimuthal and vertical
directions of the global disk, respectively. The shearing box fluid
equations read 
\begin{align} 
  &\frac{\p\rho}{\p t}+\nabla\cdot(\rho\bm{v}) = 0,\\
  &\frac{\p \bm{v}}{\p t} + \bm{v}\cdot\nabla\bm{v} +
  2\Omega_0\hat{\bm{z}}\times\bm{v} = - \frac{1}{\rho}\nabla\Pi +
  \frac{1}{\rho\mu_0}\bm{B}\cdot\nabla\bm{B}
  -\nabla\Phi,\\
  &\frac{\p\bm{B}}{\p t}= \nabla\times\left(\bm{v}\times\bm{B} -
  \eta\nabla\times\bm{B}\right), 
\end{align}
where $\rho$ is the density field; $\bm{v}$ is the total velocity in
the local frame; $\bm{B}$ is the magnetic field which satisfies
$\nabla\cdot~\bm{B}=0$; $\Pi \equiv P +
|\bm{B}|^2/2\mu_0$ is the total pressure, and $\mu_0$ is the vacuum
permeability. We choose a barotropic equation of state, specified
below, so that the gas pressure is given by $P=P(\rho)$. The
resistivity $\eta$ is either uniform or a prescribed function of
height.

The total potential is $\Phi = \Phi_\mathrm{ext} + \Phi_d$, where
\begin{align}
  \Phi_\mathrm{ext}(x,z) = -\Omega_0 S_0 x^2 +
  \frac{1}{2}\Omega_{z0}^2z^2 
\end{align}
is the effective external potential (central plus centrifugal) in the
shearing box approximation, where $S_0\equiv ~S(r_0,0)$ and
$\Omega_{z0}\equiv\Omega_z(r_0,0)$; 
and the gas potential $\Phi_d$ satisfies Poisson's equation
\begin{align}
  \nabla^2\Phi_d = 4\pi G \rho, 
\end{align}
where $G$ is the gravitational constant. For clarity, hereafter we
drop the subscript $0$ on the frequencies.

\subsection{Equilibrium disk} 
The unperturbed disk is steady and described by
$\rho=~\rho(z)$, $\bm{B} = B_z\hat{\bm{z}} + B_y\hat{\bm{y}}$ where
$B_{y,z}$ are constants and the toroidal field strength is
$B_y=\epsilon B_z$. The equilibrium 
velocity field is $\bm{v} = -Sx\hat{\bm{y}}$. We consider Keplerian
disks so that $S = 3\Omega/2$ and the epicycle frequency 
$\kappa\equiv\sqrt{2\Omega(2\Omega-S)}=\Omega=\Omega_z$.  
We assume a thin disk and neglect the radial component of the
self-gravitational force in the unperturbed disk.    

The equilibrium density field is obtained by solving
\begin{align}
  &0=\frac{1}{\rho}\frac{d P}{dz} + \Omega_z^2z + \frac{d\Phi_d}{dz},\label{eqm_eqns1}\\
  &\frac{d^2\Phi_d}{dz^2} = 4\pi G \rho.\label{eqm_eqns2}
\end{align}
We consider \begin{inparaenum}[(i)]
\item isothermal disks with $P=c_{s0}^2\rho$\label{iso_eos}; 
\item polytropic disks with $P=K\rho^2$ with $K=c_{s0}^2/2\rho_0$;
\end{inparaenum}
where $\rho_0\equiv\rho(0)$ is the midplane density.
The sound speed $c_s\equiv\sqrt{dP/d\rho}$ 
so that $c_{s0}$ is the global sound speed in the isothermal disk, and
is the midplane sound speed in the polytropic disk. For the polytropic
disk the disk thickness $H$ is such that $\rho(H)=0$. Since the
isothermal disk has no surface, we define $H$ such that
$\rho(H)=10^{-2}\rho_0$. A non-dimensional
measure of the disk thickness is given by 
\begin{align}
  f^{-1} = \frac{H\Omega}{\csmid}, 
\end{align}
and $f$ will appear in subsequent discussions.

\begin{figure}
  \includegraphics[width=\linewidth,clip=true,trim=0cm 1.5cm 0cm
    0cm]{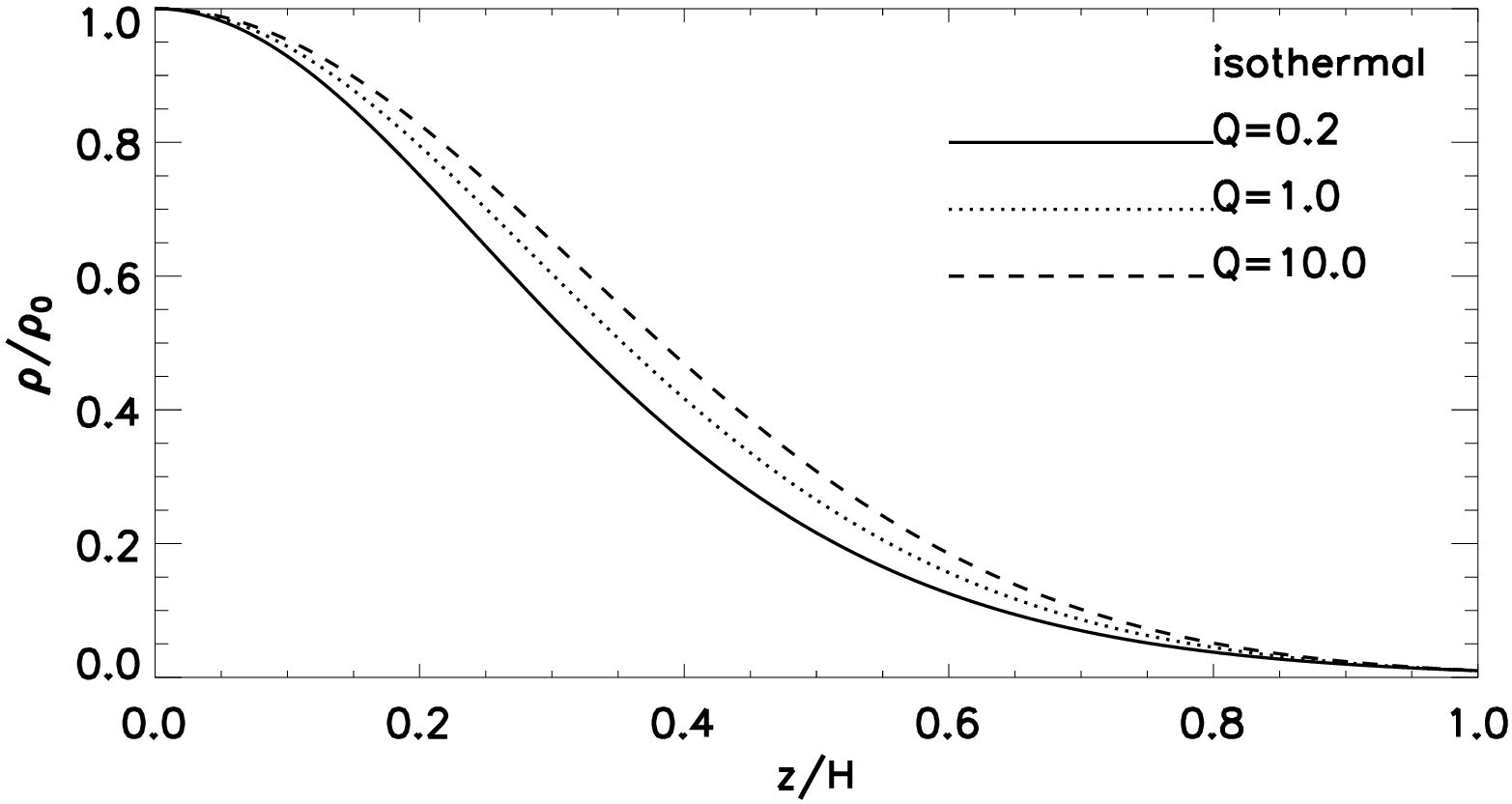} 
  \includegraphics[width=\linewidth,clip=true,trim=0cm 0cm 0cm
    0.9cm]{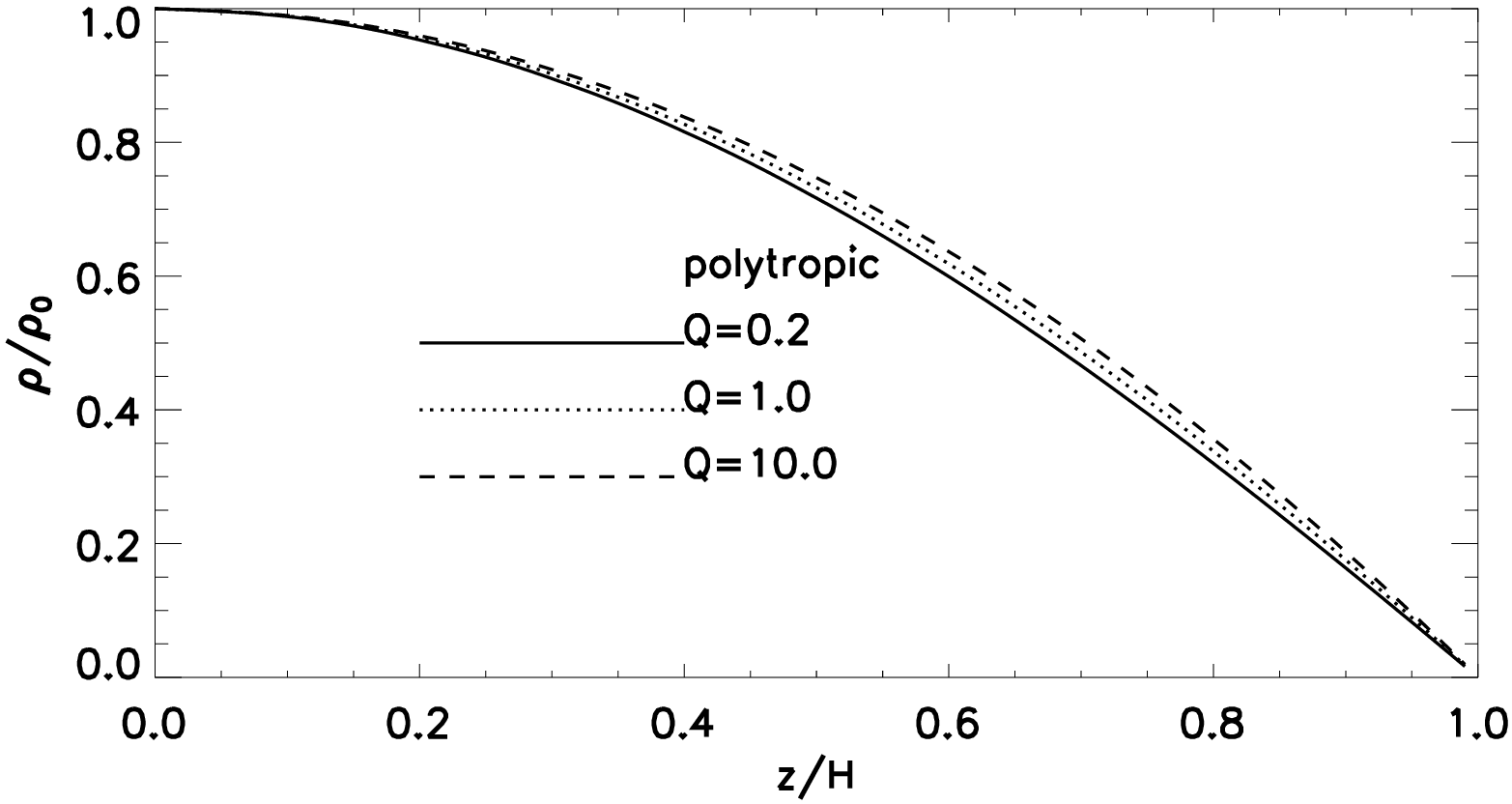} 
  \caption{Equilibrium density field from solving Eq. \ref{eqm_eqns1}
    --- \ref{eqm_eqns2} subject to an isothermal (top) and polytropic
    (bottom) equation of state. Note that the normalization for the
    horizontal axis also depends on the strength of self-gravity,
    i.e. $H=H(Q)$ and is an increasing function of $Q$. 
    \label{eqm_den}}
\end{figure}

We solve for $\hat{\rho}\equiv\rho/\rho_0$ with 
boundary conditions $\hat{\rho}=1$ and $d\hat{\rho}/dz=0$  at $z=0$. This
is done numerically for isothermal disks and analytically for the
polytropic disk (see Appendix \ref{appen1}). Examples of density
profiles are shown in Fig. \ref{eqm_den}. The normalized density field
is weakly dependent on the strength of self-gravity provided the
$z$-axis is appropriately scaled.

\subsection{Resistivity profile}\label{resis_profile}
We adopt constant resistivity or a 
resistivity prescription such that $\eta(z)$ increase towards the
midplane. In the latter case, we follow \cite{fleming03} and use 
the resistivity profile 
\begin{align}
  \eta(z) =
  \sqrt{2}\eta_0\left[\exp{\left(-g_+\right)}+\exp{\left(-g_-\right)}\right]^{-1/2},  
\end{align}
where
\begin{align}
  &g_\pm(z) =  \frac{\Sigma_\pm(z)-\Sigma_0}{\Sigma_*}, \\
  &\Sigma_\pm(z) = \int_{\pm z}^\infty\rho(z^\prime)dz^\prime, \label{sigma_pm}
\end{align}
and $\Sigma_0\equiv\Sigma_{\pm}(0)$, so that $g_\pm(0)=0$ and $\eta_0 
= \eta(0)$. The constant $\Sigma_*$ is chosen such that 
\begin{align}
  \cosh{\left(\frac{\Sigma_0}{\Sigma_*}\right)} =
  \left[\frac{\eta_0}{\eta(\infty)}\right]^2,
\end{align}
and we define $\eta_0/\eta(\infty)\equiv A$ as the conductivity 
boost factor from the midplane to the disk surface. We remark that
once $\rho$ and $d\rho/dz$ are obtained from
Eq. \ref{eqm_eqns1} --- \ref{eqm_eqns2}, the integration for
Eq. \ref{sigma_pm} can be performed implicitly by using Poisson's 
equation. 

We use the Elsasser number $\Lambda$ a non-dimensional measure of
conductivity,
\begin{align} 
  \Lambda \equiv \frac{v_A^2}{\eta\Omega},
\end{align}
where $v_A \equiv B_z/\sqrt{\mu_0\rho}$ is the vertical Alfven speed. 
Because of the density stratification, the Elsasser number 
increases with height even for constant resistivity. The disk may be
considered ideal where $\Lambda \gtrsim 1$.

\subsection{Disk parameters}
%The main parameters describing the resistive, self-gravitating
%shearing box is as follows. 
The strength of self-gravity is
parametrized by 
\begin{align}
  Q \equiv \frac{\Omega^2}{4\pi G\rho_0}
\end{align}
\citep{mamat10}, which is used to set the midplane density $\rho_0$. 
A relation between $Q$ and the Toomre parameter for gravitational
instability of razor-thin disks, $Q_\mathrm{2D}$, is described in
Appendix \ref{q3d2d}.  

The plasma $\beta$ measures the inverse strength of the  
magnetic field 
\begin{align}
  \beta \equiv \frac{\csmid^2}{v_{A0}^2} =
  \frac{\csmid^2\mu_0\rho_0}{B_z^2},  
\end{align}
where $v_{A0}$ is the midplane Alfven speed. Note that we use 
the vertical field for this definition throughout this paper. 

The strength of conductivity is measured by the midplane Elsasser
number   
\begin{align}
  \Lambda_0 \equiv\Lambda(0) =  \frac{v_{A0}^2}{\eta_0\Omega}. 
\end{align}
For non-uniform resistivity we also specify $A > 1$. 

\section{Linear problem}\label{linear}
We consider axisymmetric Eulerian perturbations to the above
equilibrium in the form $\real[\dd\rho(z)\exp{\imgi(k_xx + \sigma
    t)}]$ and similarly for other fluid variables. Here, $k_x$ is a
constant radial wavenumber and $\sigma=-(\omega +\imgi\gamma)$ is a
complex  frequency, where $-\omega$ is the real mode frequency and $\gamma$ is
the growth rate. We take $k_x>0$ without loss of generality. 
Hereafter, we suppress the exponential factor and the real part notation. 

The linearized continuity equation is 
\begin{align}
  \frac{\imgi\sigma}{c_s^2}W + \imgi k_x \dd v_x
  +\left(\ln\rho\right)^\prime\dd v_z + \dd v_z^\prime = 0,\label{lin_cont}
\end{align}
where $^\prime$ denotes $d/dz$ and $W=\delta P/\rho =
c_s^2\delta\rho/\rho$ is the enthalpy perturbation.  
The linearized equations of motion are
\begin{align}
  &\imgi\sigma \dvx - 2\Omega\dvy  = - \imgi
  k_x\w + \frac{B_z}{\mu_0\rho} \left[\dbx^\prime - \imgi 
    k_x\left(\dbz + \epsilon\dby\right)\right],\\
  &\imgi\sigma\dvy + 
    \frac{\kappa^2}{2\Omega}\dvx = \frac{B_z}{\mu_0\rho}\dby^\prime\label{lin_vy},\\ 
  & \imgi\sigma\dvz = -\w^\prime - \frac{B_y}{\mu_0\rho}\dby^\prime\label{lin_vz},  
\end{align}
where the effective enthalpy perturbation
$\w = W + \dphi$. The components of the linearized induction equation
are 
\begin{align}
&  \imgi\bar{\sigma}\dbx = B_z\dvx^\prime +
  \eta\dbx^{\prime\prime}+\eta^\prime\dbx^\prime - \imgi k_x
  \eta^\prime \dbz,\label{induct_x}\\
&\imgi\bar{\sigma}\dby = B_z\dvy^\prime -B_y\Delta - S\dbx +
  \eta\dby^{\prime\prime}+\eta^\prime\dby^\prime,\label{induct_y}\\
& \imgi\bar{\sigma}\dbz = -\imgi  k_xB_z \dvx +
\eta\dbz^{\prime\prime}, \label{induct_vert}
%& \imgi\bar{\sigma}\dbz = -\imgi  k_x \dvx  -\imgi k_x\eta \dbx^{\prime}, \label{induct_vert}
\end{align} 
where $\imgi\bar{\sigma} = \imgi\sigma + \eta k_x^2$,
$\Delta\equiv \nabla\cdot\bm{\delta v}  = \imgi k_x\dvx + \dvz^\prime$,   
and the divergence-free condition is $\imgi k_x\dbx +
\dbz^\prime=0$. Finally, the linearized Poisson equation is 
\begin{align}
  \dphi^{\prime\prime} - k_x^2\dphi = \frac{\Omega^2\rho}{c_s^2Q\rho_0}W.  \label{lin_poisson}
\end{align}

We eliminate $\dd\bm{B}$ and $\dvz$ between the linearized
equations to obtain a system of ordinary differential equations for
$\bm{U}=\left(\dvx,\dvy,W,\dphi\right)$. We detail the steps 
in Appendix \ref{reduction} for two cases considered in this paper:
%We consider two cases:
\begin{enumerate}
\item Purely vertical field with constant or variable
  resistivity, so that $\epsilon = 0$ and $\eta=\eta(z)$. 
\item Tilted field with uniform resistivity
  so that $\epsilon \neq 0$ and $\eta=\rm{constant}$.   
\end{enumerate}
Schematically, the numerical problem is to solve 
\begin{align}
  L_{11}\dvx + L_{12}\dvy + L_{13}W + L_{14}\dphi &= 0, \label{lin1}\\
  L_{21}\dvx + L_{22}\dvy + L_{23}W + L_{24}\dphi &= 0, \label{lin2}\\
  L_{31}\dvx +L_{32}\dvy + L_{33}W + L_{34}\dphi &=0,\label{lin3}\\
  \phantom{L_{31}\dvx + L_{22}\dvy +} L_{43}W + L_{44}\dphi
  &=0\label{lin4},
\end{align}
where the differential operators $L_{1j},\,L_{2j}$ and $L_{3j}$
($j=1,2,3,4$) can be read off Appendix \ref{reduction} and $L_{4j}$
($j=3,4$) from the linearized Poisson equation above. We remark that
the case of a tilted field and variable resistivity
%($\epsilon\neq0$ and $\eta=\eta(z)$
can also be reduced to the above form.

%where the differential operators $L_{1j}$, $L_{2j}$ and $L_{3j}$ can
%be read off Eq. \ref{final_vx}, Eq. \ref{final_vy} and 
%Eq. \ref{final_w} respectively, and $L_{4j}$ from
%Eq. \ref{lin_poisson}. 

%Note that $L_{13},\,L_{14},\,L_{23},\,L_{24}$ and $L_{31}$ are
%proportional to $k_x$. Then for for $k_x=0$ 
%Eq. \ref{lin1}---\ref{lin2} are decoupled from 
%Eq. \ref{lin3}---\ref{lin4}. In this case, a solution is $\dphi=W=0$,
%with $\dvx$ and $\dvy$ determined by $L_{11}\dvx + L_{12}\dvy = 0$ and 
%$L_{21}\dvx + L_{22}\dvy=0$, which is the incompressible MRI problem. 
%This means that the density and potential perturbations can only
%influence the MRI for $k_x\neq0$. Nevertheless, for $k_x=0$,
%self-gravity can still have an effect through the background
%equilibrium.  

\subsection{Domain and boundary conditions}\label{domain}
 
For a vertical field, considered in \S\ref{result1} and
\S\ref{result2}, we take $\bm{U}$ to be an even function of
$z$. Odd modes are permitted in \S\ref{result3}, where an azimuthal field may
be included. In both
setups the gravitational potential boundary condition, given by
\cite{goldreich65a}, is
%The potential boundary
%condition is 
%\begin{align}
%  \dphi^\prime(\zmax) + k_x\dphi(\zmax) = 0 \label{pot_bc1},
%\end{align}
%and is adapted from \cite{goldreich65a}. The general form of this
%potential boundary condition is used for case 2 below.
%and 
%the potential boundary condition is 
\begin{align}
  \dphi^\prime(\pm \zmax) \pm k_x\dphi(\pm\zmax) =  - \left. \frac{\Omega^2
    \rho \xi_z}{\rho_0Q}\right|_{\pm\zmax}, \label{pot_bc2}
\end{align}
where $\xi_z=\dvz/\imgi\sigma$ is the vertical Lagrangian
displacement, and $z=\pm Z_s$ is the upper and lower disk surfaces,
respectively. 

%In practice,  Eq. \ref{pot_bc1} and \ref{pot_bc2} are
%very similar because $\rho(\pm\zmax)\ll\rho_0$. 
%}

\subsubsection{Case 1: vertical field}
 Here we impose $d\bm{U}/dz=0$ at $z=0$. This permits higher
numerical resolution by reducing the computational domain to
$z\in~[0,Z_s]$. At the upper disk boundary $z=Z_s$ we set  
\begin{align}
  \dbx(Z_s) = \dby(Z_s) = \dvz (Z_s) = 0, 
\end{align}
so the field remains vertical. The derivation of the magnetic
  field boundary conditions may be found in \cite{sano99}.

\subsubsection{Case 2: tilted field} 
In this more general setup the computational domain is
$z\in[-Z_s, Z_s]$ and no symmetry across the midplane is
  enforced. At the disk surfaces we adopt the `halo' model of
\cite{gammie94}, so that
\begin{align}
  \Delta(\pm Z_s) & = 0, \\
  \dby(\pm Z_s) & = 0,\\
  \dbz(\pm Z_s) \mp \imgi \dbx(\pm Z_s) &= 0,
\end{align}
     and this case permits $\dd v_z(\pm Z_s)\neq0$.
%      In practice, the gravitational potential boundary condition is
%      similar to case 1 because $\rho(\pm Z_s)\ll\rho_0$.}
    
\subsection{Numerical procedure}
We use a pseudo-spectral method to solve the set of linearized
equations. Let
\begin{align}\label{cheby_expand}
  \bm{U}(z) 
  = \sum_{k=1}^{N_z} \bm{U}_k\psi_k(z/\zmax), 
\end{align}
where 
\begin{align}
  \psi_k  = 
  \begin{cases}
    T_{2(k-1)} & B_y \equiv 0 \text{ (case 1)},\\
    T_{k-1}   & B_y\neq 0 \text{ (case 2)},
  \end{cases}
\end{align}
and  $T_l$ is a Chebyshev polynomial of the first kind of order $l$
\citep{stegun65}. Note that for case 1 the midplane symmetry condition
is taken care of by the choice of basis functions. 

The pseudo-spectral coefficients $\bm{U}_n$  are obtained by demanding
the set of linear equations to be satisfied at $N_z$ collocation
points along the vertical direction, here chosen to be the 
extrema of $T_{l_\mathrm{max}}$ plus end points, where 
$l_\mathrm{max}$ is the highest polynomial order. Our standard
resolution is $N_z=256$ ($N_z=257$) for case 1 (case 2).   

The above procedure discretize the linear equations to a matrix equation,
\begin{align}\label{matrix_eqn}
\bm{M}\bm{w} = \mathbf{0}, 
\end{align}
where $\bm{M}$ is a $4N_z\times 4 N_z$ matrix representing the $L_{ij}$ 
plus boundary conditions, 
and $\bm{w}$ is a vector storing the pseudo-spectral coefficients. 
Starting with an initial guess for $\sigma$, non-trivial solutions to
Eq. \ref{matrix_eqn} are obtained by varying $\sigma$ using Newton-Raphson iteration 
such that $\mathrm{det}\bm{M}=0$ \citep[details can be found in][]{lin12}.

\subsubsection{Non-dimensionalization}\label{non-dim}
We solve the linearized equations in non-dimensional form,
by defining
%We further write 
\begin{align}
  &z=\hat{z}H,\quad k_x =  \hat{k}_x/H,\quad \sigma = \hat{\sigma}\Omega,
  \quad \delta\bm{v} = \csmid 
  \delta\hat{\bm{v}}, \\ 
  &\delta\bm{B} = B_z\delta\hat{\bm{B}},\quad
  \delta\rho = \rho\hat{W}/\hat{c}_s^2,\quad \delta\Phi =
  \csmid^2\delta\hat{\Phi},  
\end{align} 
where $\hat{c}_s=c_s/\csmid$. We also non-dimensionalize 
background quantities, i.e. $\hat{v}_A=v_A/\csmid$,
$\hat{S}=S/\Omega$, $\hat{\kappa}=\kappa/\Omega$,
$\hat{\Omega}_z=\Omega_z/\Omega$ and $\hat{\eta} = \eta/(H^2\Omega)$.

\subsection{Diagnostics}
We visualize results in terms of dimensionless energy densities. We
define 
\begin{align}
  &E_m \equiv \frac{|\delta \hat{\bm{B}}|^2}{2\beta},\\
  &E_g = 
  \frac{\hat{\rho}}{2\hat{c}_s^2}\left|\real\left(\hat{W}\dd\hat{\Phi}^*\right)\right|,\\  
  &E_k = \frac{1}{2}\hat{\rho}|\delta\hat{\bm{v}}|^2,\\
  &E_t = \frac{\hat{\rho}|\hat{W}|^2}{2\hat{c}_s^2},
\end{align}
as the perturbed magnetic, gravitational, kinetic and thermal
energies, respectively, which are functions of $z$. Although we do not
solve an energy equation, we nevertheless define $E_t$ as a measure of
density perturbations \citep{kojima89}. The total energy
is $E=E_m+E_g+E_k+E_t$. We use $\avg{\cdot}$ to denote an average over
$z$. 

Since we will primarily be concerned with massive disks, we define
\begin{align}
  \tau \equiv \frac{\avg{E_g}}{\avg{E_g} + \avg{E_m}} 
\end{align} 
as a measure of the importance of self-gravity. Thus, modes with $\tau
= 1$ are energetically dominated by self-gravity (GI) and modes with
$\tau\ll1$ are dominated by magnetic perturbations (MRI).

\section{MRI in self-gravitating disks}\label{result1}
In this section we focus on the MRI and use the vertical field setup
of case 1. We first consider MRI modes with negligible density/potential
perturbations to see the effect of self-gravity on the MRI through the
background stratification, then go on to examine MRI modes with
density/potential perturbations in massive disks.  

\subsection{Influence of self-gravity on the MRI through the
  background equilibrium}
Here we use polytropic disks, which have a well-defined disk
thickness. The upper disk boundary is set to $Z_s=0.99H$. We fix
$\beta=100$ and $k_xH=0.1$ unless otherwise stated. 

\subsubsection{Uniform
  resistivity}   
\begin{figure}
  \includegraphics[width=\linewidth]{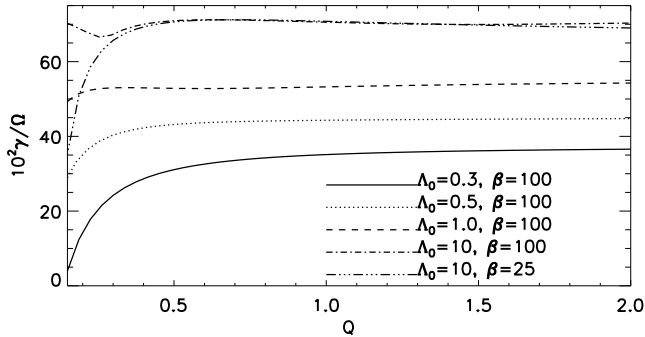}
  \caption{ MRI growth rates as a function of $Q$ and midplane 
    Elsasser numbers $\Lambda_0$, in polytropic disks with $\beta=100$
    (solid, dotted and dot-dashed lines). The dash-triple dot line is
    the $\Lambda_0=10$ case with $\beta=25$. The resistivity is
    uniform.   
    \label{compare_growth_poly_uniresis}}
\end{figure}

Fig. \ref{compare_growth_poly_uniresis} plots MRI growth rates as a
function of $Q$ and $\Lambda_0$. The resistivity is uniform ($A=1$). 
For ideal MHD and a weak field ($\Lambda_0>1$, $\beta=100$), there is
negligible dependence on $Q$. However, with $\beta=25$ or 
in the resistive limit  ($\Lambda_0<1$), growth rates decrease noticeably for $Q<0.5$
($Q_\mathrm{2D}\lesssim 1.5$).  Since we find density and potential
perturbations to be negligible (i.e. the linear response is
non-self-gravitating), this shows that disk self-gravity can affect the
MRI through the background equilibrium. 

\cite{sano99} found that for MRI to operate, its 
wavelength $\lambda$ should fit inside the disk. That is,   
\begin{align}\label{sano_crit}
  \lambda \equiv
  \mathrm{max}\left(\lambda_\mathrm{ideal},\lambda_\mathrm{resis}\right)\lesssim
  2H, 
\end{align}
where the MRI wavelengths are given by 
\begin{align}\label{lambda_ideal}
  \frac{\lambda_\mathrm{ideal}}{2H} = \frac{4\pi}{\sqrt{15}} f \hat{v}_A =
  \frac{4\pi f}{\sqrt{15\beta\hat{\rho}}}
\end{align}
for ideal MHD, and 
\begin{align}\label{lambda_resis}
  \frac{\lambda_\mathrm{resis}}{2H} = \frac{2\pi}{\sqrt{3}}\frac{\hat{\eta}}{\hat{v}_A f} =
  \frac{2\pi f}{\Lambda_0}\sqrt{\frac{\hat{\rho}}{3\beta}} 
\end{align}
in the limit of high resistivity.  
   
Because $\hat{\rho}$ is weakly dependent
on $Q$ (Fig. \ref{eqm_den}), self-gravity only affects the
MRI through the factor $f$, which increases with decreasing $Q$ (see
Fig. \ref{plot_fq} in Appendix \ref{appen1}).   
This implies that sufficiently strong self-gravity can stabilize the
MRI by making $ 2H<\lambda $.   
 
\begin{figure}
  \includegraphics[width=\linewidth]{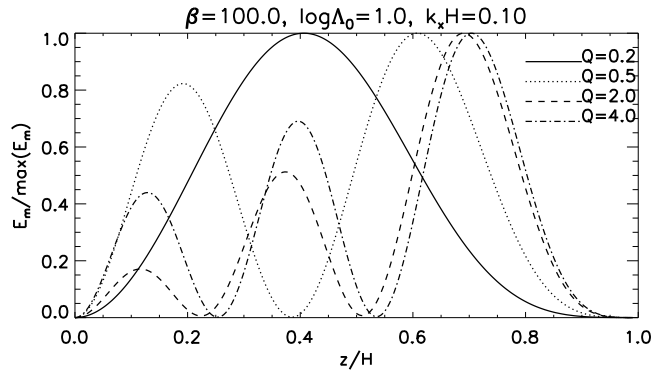}
  \caption{MRI magnetic energies in ideal polytropic disks
    for different strengths of self-gravity.    
    \label{compare_result_lambda10}}
\end{figure}

In the ideal limit with $\beta=100$, we find $\lambda < 2H$
throughout most of the disk for the values of $Q$ considered, so
self-gravity does not affect growth rates significantly. However, the
ratio $\lambda/2H$ does increase with stronger
self-gravity. Consequently, the wavelength of the instability, in
units of $H$, increases. This is shown in
Fig. \ref{compare_result_lambda10} which plots the magnetic energies
for $\Lambda_0=10$ and a range of $Q$ values. The number of vertical 
nodes decrease with $Q$, i.e. the disk accommodates fewer wavelengths
because increasing vertical self-gravity makes it thinner.

We repeated the $\Lambda_0=10$ case with a stronger field $\beta=25$,
shown in Fig. \ref{compare_growth_poly_uniresis} as the dashed-triple
dot line. Here, strong self-gravity is effective in reducing the
growth rate, because decreasing $\beta$ enhances the dependence of
$\lambda/H$ on $f(Q)$. For $Q=0.2$ and $\beta=25$ we find 
$\lambda_\mathrm{ideal}/2H\sim 1$ at the midplane and the growth rate is reduced significantly.

\begin{figure}
  \includegraphics[width=\linewidth]{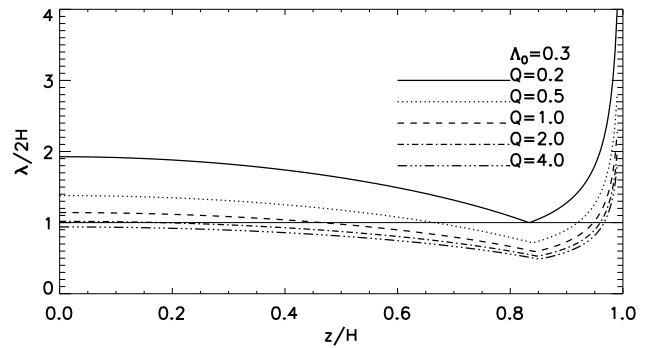}
  \caption{Approximate wavelengths of the most unstable MRI modes as given by
    Eq. \ref{sano_crit}---\ref{lambda_resis}, normalized by the 
    disk thickness, as a function of height. MRI is expected to
    operate if $\lambda/2H\lesssim 1$. 
    \label{lambda_poly_resis}}
\end{figure}

Self-gravity also appreciably decreases the MRI growth rates in the
resistive limit. Fig. \ref{lambda_poly_resis} plots 
Eq. \ref{sano_crit} for $\Lambda_0=0.3$. In the non-self-gravitating
disk ($Q=4$) the instability criterion is marginally satisfied and the
MRI operates. As $Q$ decreases, 
Eq. \ref{sano_crit} is violated and the MRI growth rate is
significantly reduced. This is seen for $Q=0.2$ where $\lambda \geq 2H$ throughout
the disk. (The instability is not suppressed since
Eq. \ref{lambda_ideal}---\ref{lambda_resis} is only exact for
unstratified disks.) Although the function 
$f(Q)$ does not change significantly for the range of $Q$ considered,
the dependence of $\lambda/H$ on $f(Q)$ is amplified by the
denominator $\Lambda_0<1$ in the resistive case. Modes in
Fig. \ref{lambda_poly_resis} have no nodes in the magnetic energy
$E_m$ except at $z\simeq0,\,H$, i.e. only the longest wavelength survives
 against large resistivity. 
%self-gravity tips it over    

\subsubsection{Layered 
  resistivity} 
\begin{figure}
  \includegraphics[width=\linewidth]{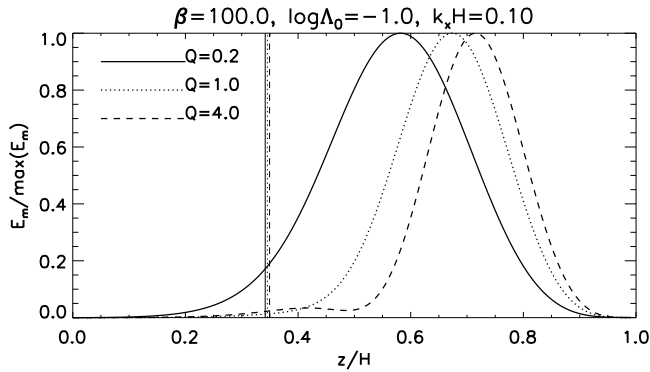}
  \caption{Magnetic energies as a function of height, for polytropic disks
    in which the conductivity increases by a
    factor $A=10^2$ in going from the midplane to the upper disk
    boundary. The vertical lines indicate $\Lambda=1$ for each value 
    of $Q$.
    \label{poly_layer}}
\end{figure}

Here we consider disks with midplane Elsasser number $\Lambda_0=0.1$
and a variable resistivity profile with
$A=10^2$. Fig. \ref{poly_layer} compares the magnetic  
energies for $Q=0.2,\,1$ and $4$. They have similar growth rates, $\gamma/\Omega
= 0.53,\,0.64$ and $0.66$, respectively. In the non-self-gravitating
limit ($Q=4$), the MRI is effectively suppressed for
$z\lesssim0.5H$. This is consistent with the picture of layered
accretion proposed for non-self-gravitating disks \citep{gammie96,fleming03}. 
However,in the massive disk ($Q=0.2$) the mode occupies a wider vertical
extent because its wavelength (in units of $H$) is larger. This
suggests that in massive disks, the MRI is not well localized to a
sub-layer within the height, even when the resistivity has a layered
structure.

%We find qualitatively similar results for large $A$ with small
%  $\Lambda_0$. 
We also performed additional calculations with $A=10^3$ and $A=10^4$
\citep[see Fig. 1 of ][]{gressel12}. For $\Lambda_0=0.1$, we find no
significant increase in the magnetic energy in the resistive zones. 
However, lowering $\Lambda_0$ gives similar
results to Fig. \ref{poly_layer}, e.g. for $A=10^3$ and
$\Lambda_0=10^{-2}$ or $A=10^4$ and $\Lambda_0=10^{-3}$ the magnetic
energy penetrates into the resistive zone for strongly
self-gravitating disks. In general, the magnetic energy density maximum moves
toward the midplane with increasing self-gravity.

\subsubsection{Dependence on $k_x$}
\begin{figure}
  \includegraphics[width=\linewidth]{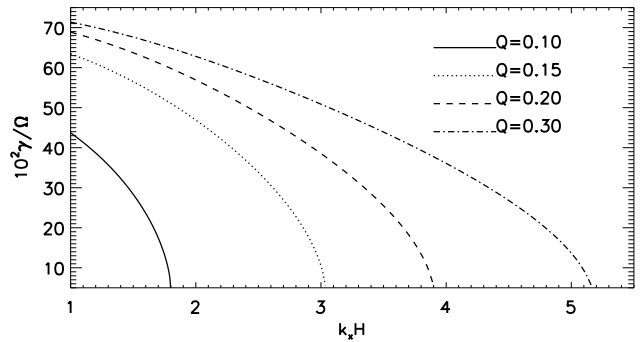}
  \caption{MRI growth rates in self-gravitating polytropic disks, as a
    function of the horizontal wavenumber $k_x$. The disk is ideal
    ($\Lambda_0=10^2,\, A=1$) with $\beta = 40$. These modes have negligible
    density/potential perturbations.  
    \label{compare_growth_poly_kx}}
\end{figure}

The above experiments show that with increasing 
disk self-gravity, the MRI becomes more global in the vertical
direction. We find a similar result in the horizontal direction. 
Fig. \ref{compare_growth_poly_kx} show MRI growth rates as a
function of $k_x$ for a range of $Q$ values. Increasing self-gravity
decreases the cut-off radial wavenumber for the MRI. We checked
that these modes have negligible density perturbations. Then we can 
understand this result by invoking the instability criteria for  
incompressible MRI in an unstratified Keplerian disk,
\begin{align}
  v_A^2(k_z^2 + k_x^2) < 3\Omega^2,
\end{align}
where $k_z$ is a vertical wavenumber \citep{kim00}. Setting $k_z^2\sim
\Omega^2/v_A^2$ and non-dimensionalizing, we find
\begin{align} 
  k_xH \lesssim \frac{\sqrt{\beta}}{f},
\end{align}
where order-unity factors have been dropped. Despite a simplistic
approach, this demonstrates that with increasing self-gravity
(increasing $f$), we expect MRI modes with small radial
length scales to be suppressed.

\subsection{Influence of self-gravity on the MRI through the linear
  response}  
Our goal here is to examine whether or not self-gravity can amplify
the density perturbations associated with the MRI. We compute unstable modes
in a massive isothermal disk with $Q=0.2$ (corresponding to
$Q_\mathrm{2D}=0.72$), which is still expected to be marginally  
stable to gravitational instability \citep[][who find
a critical value of  $Q\simeq 0.2$]{mamat10}.   
The upper disk boundary is set to $Z_s=H$. 

%resolution Nz=256

\subsubsection{Ideal disks}
\begin{figure}
  \includegraphics[width=\linewidth]{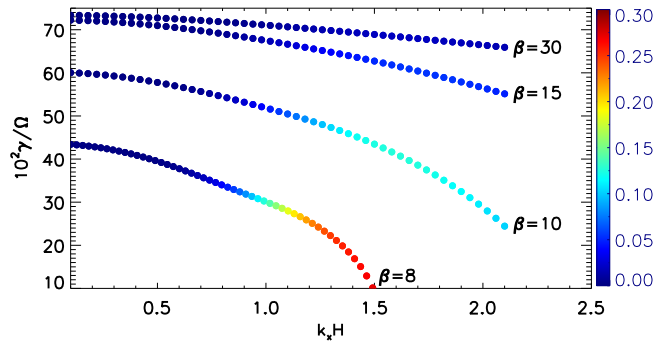}
  \caption{Growth rates of MRI modes in isothermal self-gravitating
    disks with $Q=0.2$ ($Q_\mathrm{2D}=0.72$) in the limit of ideal MHD
    ($\Lambda_0=10^2$, $A=1$), for a range of field strengths $\beta$. The
    colorbar measures the importance of self-gravity by $\tau$.  
    %{\bf ADD BETA=8}
    \label{gravity_energy}}
\end{figure}

We first consider ideal MHD by adopting a uniform 
resistivity with $\Lambda_0=100$.  
Fig. \ref{gravity_energy} plots MRI growth rates as a function of
$k_x$ for several values of $\beta$. The curves are color-coded 
according $\tau$. (Recall $\tau\to1$ implies self-gravity dominates
over magnetic perturbations, and $\tau\to0$ is the opposite limit.)  
The potential perturbation is negligible for all cases when
$k_xH\lesssim 0.5$, since the MRI becomes incompressible as $k_x\to
0$.  

For $\beta\gg 1$, i.e. a weak field, density perturbations are
negligible and the incompressible MRI operates. However, as $\beta$ is
lowered and the MRI growth rate reduced, we find non-negligible
potential perturbation for $k_xH=O(1)$. This suggests that in a
strongly magnetized disk that still permits the MRI, the associated
density perturbation can be important when the disk is
self-gravitating. %but formall stable 

\subsubsection{Resistive disks}
\begin{figure}
  \includegraphics[width=\linewidth]{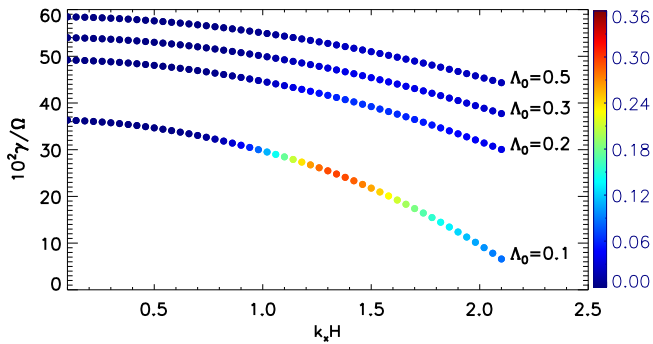}
  \caption{
    Growth rates of MRI modes in an isothermal self-gravitating
    disk with $Q=0.2$ ($Q_\mathrm{2D}=0.72$) at fixed $\beta=100$, 
    for a range of midplane Elsasser numbers. The resistivity is
    uniform. The colorbar measures the importance of self-gravity by $\tau$. 
    \label{gravity_energy_resis}}
\end{figure}

\begin{figure}
  \includegraphics[width=\linewidth]{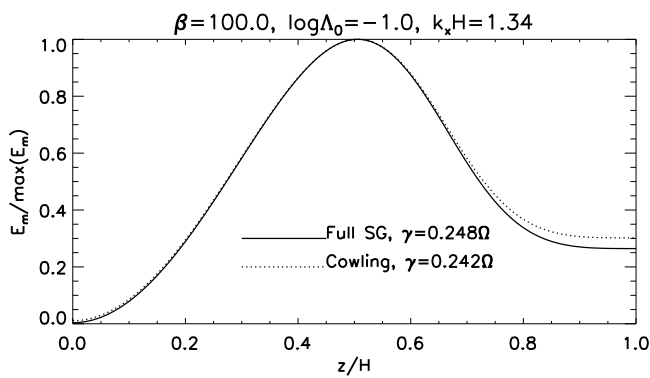}
  \caption{Magnetic energy associated with the linear mode with
    largest gravitational-to-magnetic energy ratio in 
    Fig. \ref{gravity_energy_resis} (solid) compared with that computed
    under the Cowling approximation (dotted). %{\bf REDO FIGURE}
    \label{mri_massive_cowling}}
\end{figure}

We repeat the above calculation for resistive disks, but fix
$\beta=100$ and vary the midplane Elsasser number $\Lambda_0$. Growth
rates are shown in Fig. \ref{gravity_energy_resis}. 
Interestingly, the highly resistive case $\Lambda_0=0.1$ has
comparable magnetic and gravitational energies: at $k_xH\simeq1.3$ we
find $\tau\sim 0.3$, which corresponds to $\avg{E_g}\sim 0.5\avg{E_m}$.  
Fig. \ref{mri_massive_cowling} compares the magnetic
energy of this mode to that computed in the \emph{Cowling  
  approximation}, where the Poisson equation is ignored in the 
linearized equations and the potential perturbation set to zero  
(formally letting $Q\to\infty$ in Eq. \ref{lin_poisson}).  
The growth rate increases when self-gravity is included in the linear
response, since self-gravity is usually destabilizing. However, 
$\gamma$ and $E_m$ are very similar, indicating that the instability
in the self-gravitating calculation is fundamentally still the MRI.   

\begin{figure}
  \includegraphics[width=\linewidth]{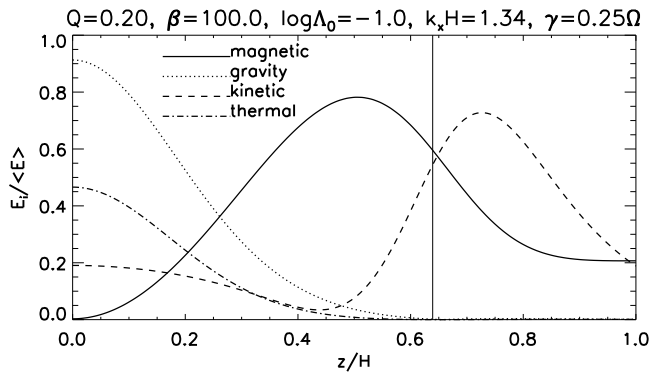}
  \caption{Example of a resistive MRI mode with significant
    gravitational potential perturbation. The disk is isothermal. 
    The vertical line
    indicates $\Lambda=1$.  
    \label{mri_massive_resis}}
\end{figure}

Fig. \ref{mri_massive_resis} plots the energies associated with the 
MRI mode discussed above. 
The gravitational energy exceeds the 
magnetic energy near the midplane ($z\lesssim0.2H$). The growth rate
$\gamma=0.25\Omega$ is not much smaller than that of the most unstable
mode ($\gamma=0.36\Omega$ for $k_xH=0.1$), so significant density
perturbations will grow on dynamical timescales for this
system, even though GI is not expected.

\subsubsection{Qualitative interpretation} 
To make sense of the above results, we first return to ideal MHD and 
consider regions close to the disk midplane ($z\sim 0$), where
self-gravity is expected to be most important. For this discussion we
will ignore stratification and set $d^2/dz^2\to -k_z^2$. The governing
equations are then 

\begin{align}
  &  0= v_A^2k^2\dvx + \imgi\sigma\left(\imgi\sigma \dvx - 2\Omega\dvy + \imgi k_x \w\right),\label{simp1}\\
  &  0= v_A^2k_z^2\left(\dvy + \frac{\imgi S
  }{\sigma}\dvx\right) + \imgi\sigma\left(\imgi\sigma\dvy +
  \frac{\kappa^2}{2\Omega}\dvx\right),\label{simp2}\\
  & 0 = -k_z^2\w + \frac{\sigma^2}{c_s^2}W + \sigma k_x \dvx, \label{simp3}\\
  & 0 = k^2\dphi + \frac{\Omega^2}{c_s^2Q}W,\label{simp4}
\end{align}
where $k^2 = k_z^2 + k_x^2$. We imagine an iterative procedure to
solve the above equations, starting from the Cowling approximation
where $\delta\Phi\to0$ and $Q\to\infty$. This is the standard MRI and we denote the
solution as $\dvx^{(0)}$, $\dvy^{(0)}$ and $W^{(0)}$. Eq. \ref{simp3} implies
\begin{align}
  W^{(0)} = \frac{c_s^2\sigma k_x \dvx^{(0)}}{c_s^2k_z^2 - \sigma^2}. 
\end{align}
We argue below that $c_s^2k_z^2\gg \sigma^2$ by taking $k_z\sim
\Omega/v_A$. Then, recalling $W=c_s^2\delta\rho/\rho$, we can write
\begin{align}
  \frac{\delta\rho^{(0)}}{\rho} \sim
  \frac{\sigma}{\Omega}\frac{1}{\beta}\left[\frac{k_x\dvx^{(0)}}{\Omega}\right].   
\end{align}
The MRI has, in general, a non-zero density perturbation. However, it
is negligible for $k_x\to 0$ and/or a weak field ($\beta \gg 1$). 

% In general, the
%compressible MRI has $W^{(0)}\neq0$, since

We now include self-gravity.  The Poisson equation implies $W^{(0)}$
has an associated potential perturbation,     
\begin{align} 
  \dphi = -\frac{\Omega^2}{c_s^2 Q k^2} W^{(0)}.
\end{align}
Physically, we expect $k^2\geq0$, so that a positive (negative) local density
perturbation causes a negative (positive) local potential
perturbation. We then insert $\dphi$ back into the momentum and
continuity equations, and ask how does this potential perturbation
modify the Cowling solution? Writing $\dvx^{(0)} \to \dvx^{(0)} +
\dvx^{(1)}$ and similarly for $\dvy $ and $ W$, we find

\begin{align}
  &   k_x\sigma \dphi = v_A^2k^2\dvx^{(1)} + \imgi\sigma\left[\imgi\sigma
  \dvx^{(1)} - 2\Omega\dvy^{(1)} + \imgi k_x W^{(1)}\right], \label{simp_pert1}\\ 
  &  0= v_A^2k_z^2\left[\dvy^{(1)} + \frac{\imgi S
    }{\sigma}\dvx^{(1)}\right] + \imgi\sigma\left[\imgi\sigma\dvy^{(1)} +
  \frac{\kappa^2}{2\Omega}\dvx^{(1)}\right],\label{simp_pert2}\\
  & k_z^2\dphi  = \left(\frac{\sigma^2}{c_s^2}-k_z^2\right)W^{(1)} +
  \sigma k_x \dvx^{(1)}.\label{simp_pert3} 
\end{align}
Now, if the perturbations to the magnetic field remain
unchanged, i.e. the mode remains close to the standard MRI as observed
in Fig. \ref{mri_massive_cowling}, then $\dvx^{(1)} \sim 0$ and
$\dvy^{(1)}\sim0$, so Eq. \ref{simp_pert2} is 
satisfied. Eq. \ref{simp_pert1} then require $\dphi +
W^{(1)}\sim0$. This is compatible with Eq. \ref{simp_pert3} if 
\begin{align}
  \left|k_z^2\right| \gg \left|\frac{\sigma^2}{c_s^2}\right|. \label{cond}
\end{align}
For the ideal MRI,  we take $k_z\sim \Omega/v_A$. Then
$|\sigma^2/c_s^2k_z^2|\sim |\sigma^2/\Omega^2\beta|\ll1$ because
$|\sigma|\lesssim \Omega$ and we are considering $\beta\gtrsim 10$. 
Thus Eq. \ref{cond} is generally satisfied.   

The above assumptions imply
\begin{align}
  W^{(1)} \sim \frac{\Omega^2}{c_s^2 Q k^2} W^{(0)},\label{feedback}
\end{align} 
which indicates a non-zero density perturbation due to the
MRI can be amplified by self-gravity. Now, for $k_xH\sim 1$ we have
$|k_z^2/k_x^2|\sim \beta/f^2\gg1$ because $f=O(1)$ and $\beta\gtrsim10$
for the cases considered above. Then
\begin{align}
  \left|\frac{W^{(1)}}{W^{(0)}}\right| \sim \frac{1}{Q\beta}, 
\end{align}
 suggesting stronger amplification of the density field by 
 self-gravity with increasing field strength (decreasing
 $\beta$).

 The above arguments can be adapted to the resistive
 disk. Eq. \ref{simp3}---\ref{simp4} are unchanged, while resistive
 terms appearing in Eq. \ref{simp1}---\ref{simp2} only involve the
 potential perturbation through $\w$. For the resistive MRI we take
 $k_z\sim v_A/\eta$ and $|\sigma|\sim v_A^2/\eta = \Lambda\Omega$
 \citep{sano99}. Then $|\sigma^2/c_s^2k_z^2|\sim 1/\beta \ll 1$ so
 Eq. \ref{cond} is satisfied. Noting that $k_z^2\sim
 \Lambda^2\Omega^2\beta/c_s^2$, the feedback equation becomes
 \begin{align}
   \left|\frac{W^{(1)}}{W^{(0)}}\right| \sim
   \frac{1}{Q\left(f^2\hat{k}_x^2 + \beta\Lambda^2\right)},
 \end{align}
 so increasing the resistivity (decreasing $\Lambda$) should enhance
 density perturbations.  

%these arguments assume, of course, there is a non-zero density
%perturbation in the first place for self-gravity to ehance. 

 For weak fields in an ideal disk, the MRI has a vertical
 wavelength $\lambda \ll H$. It will be almost incompressible so the 
 `seed' density perturbation $W^{(0)}$ is small.  
 The perturbed mass contained within 
 $\sim \lambda$ is small and its potential is
 unimportant. Furthermore, considering the stratified  
 disk, $\lambda\ll H$ imply rapid variations in the density
 perturbation across the disk height, averaging to zero, so the
 magnitude of the associated potential perturbation is small. 
 Self-gravity does not affect the MRI in this regime.   
 
%the `seed' density pert is also small
%weak amplification

 However, a strong field and/or large resistivity increases the MRI
 vertical wavelength. When the vertical scale of the MRI becomes
 comparable to the disk thickness, i.e. $\lambda\sim H$, the
 perturbed mass across the disk height can contribute to a net potential
 perturbation. We therefore expect a necessary condition for
 self-gravity to affect the MRI is for the latter to be weak. 
%we are not considering GI yet

%Our
% interpretation above also assumes that the Cowling solution can
% provide some `seed' density perturbation. 
%Thus we also require the
% MRI to be compressible. 
%there has to be some  `seed' density perturbation provided by the
%non-SG MRI 

\section{Gravitationally unstable disks} \label{result2}
Gravitational instability becomes possible in a sufficiently
massive and/or cold disk. Here, we explore whether or not GI and MRI
can interact by computing unstable modes for isothermal disks with $Q < 0.2 $
($Q_\mathrm{2D}\lesssim ~0.67$) which permits GI, as shown below.
We consider ideal disks with $\Lambda_0=100$ and $A=1$, unless
otherwise stated.  

\subsection{Co-existence of MRI and GI} 
\begin{figure}
  \includegraphics[width=\linewidth,clip=true,trim=0cm 2.1cm 0cm
    0cm]{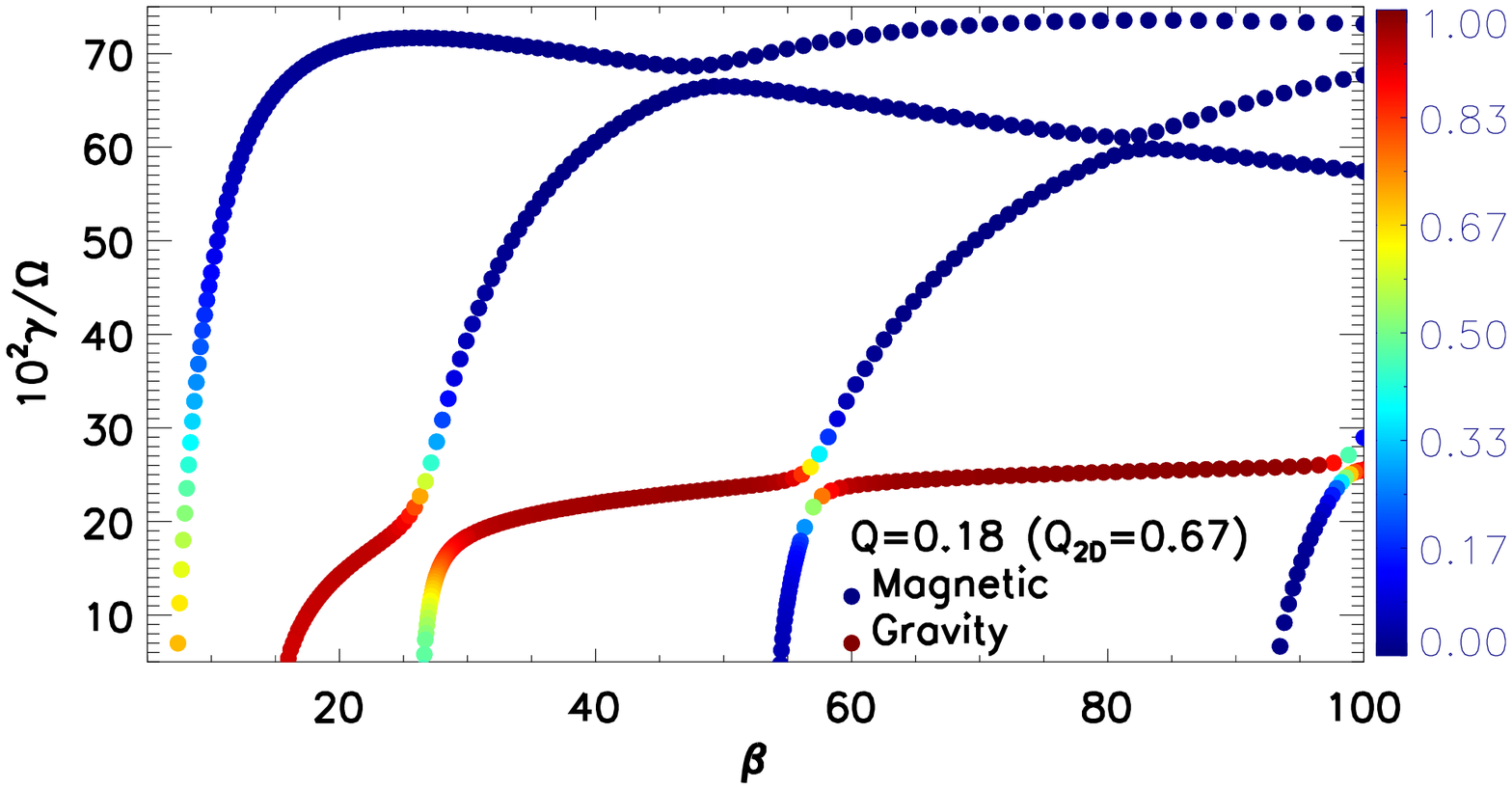} 
  \includegraphics[width=\linewidth,clip=true,trim=0cm 2.1cm 0cm
    0.5cm]{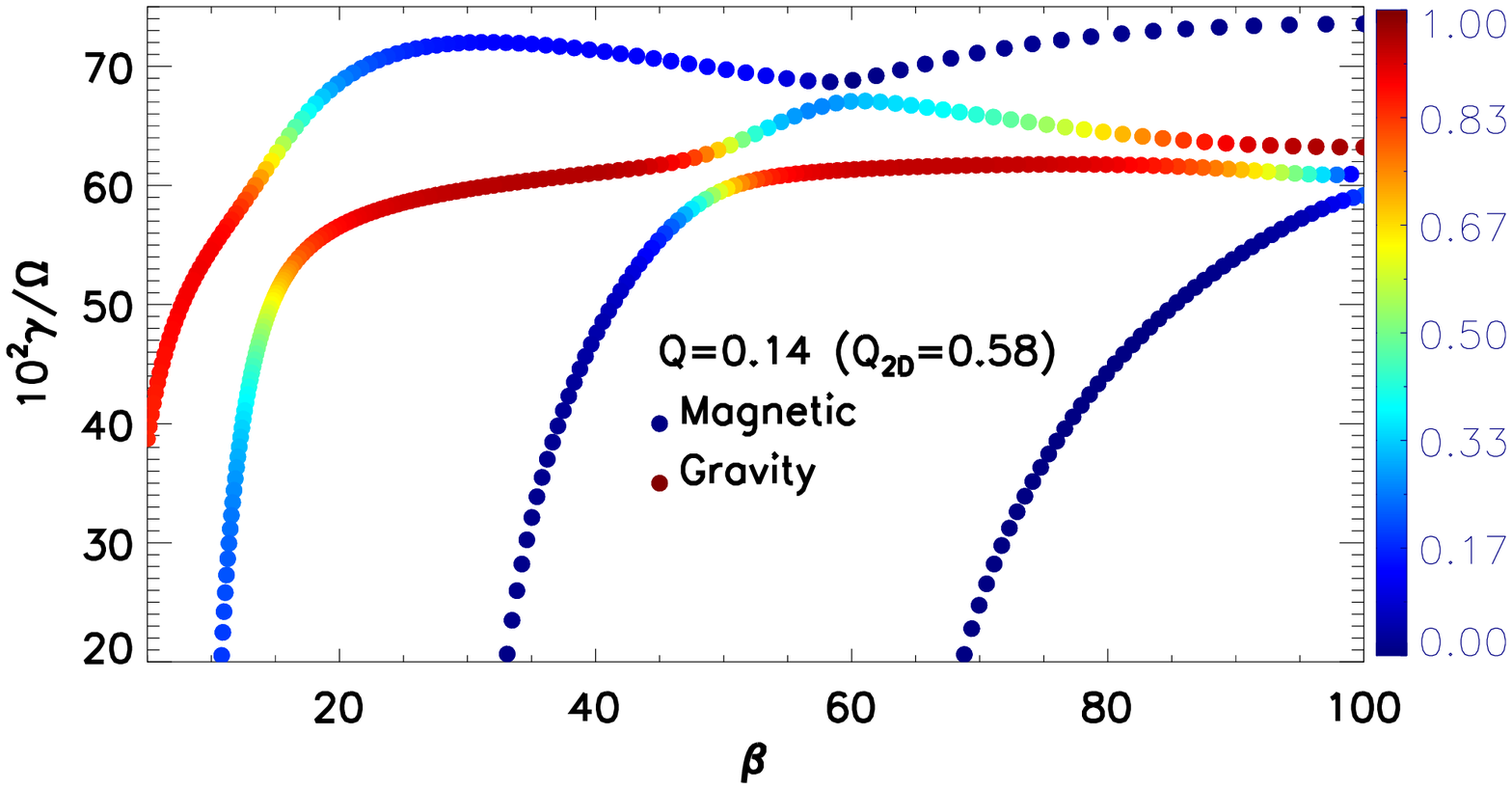} 
  \includegraphics[width=\linewidth,clip=true,trim=0cm 0cm 0cm
    0.5cm]{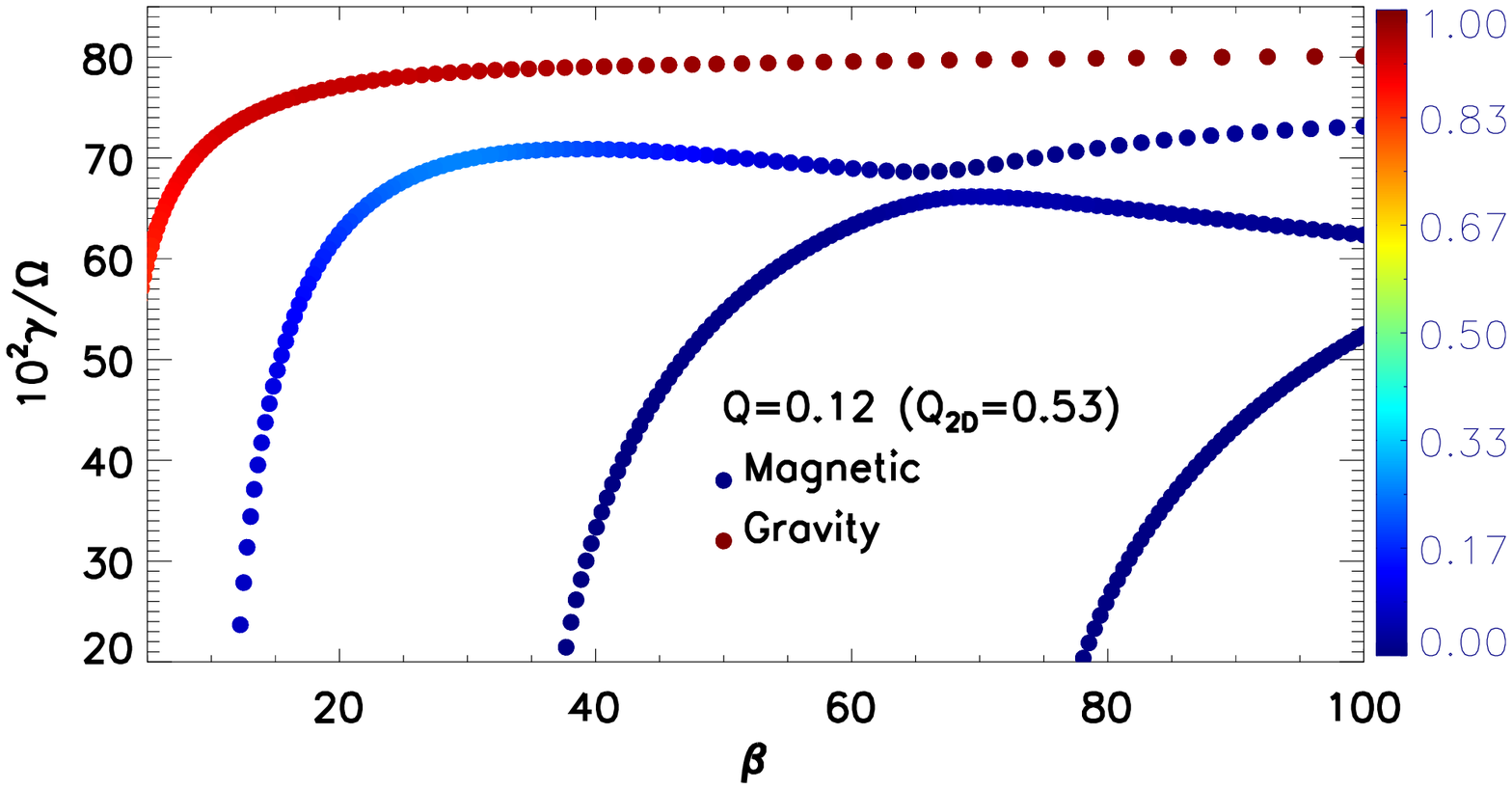} 
  \caption{Growth rates for modes with $k_xH=1$ in isothermal ideal
    disks with $Q=0.18$ (top), $Q=0.14$ (middle) and $Q=0.12$ (bottom). The
    colorbar measures the importance of  
    self-gravity by $\tau$.   
    \label{compare_growth3}}
\end{figure}

Fig. \ref{compare_growth3} show growth rates for modes with $k_xH=1$
as a function of $\beta$ in disks with $Q=0.18$, $Q=0.14$ and
$Q=0.12$.   
All three cases display distinct GI modes (red/brown branch). The GI 
growth rates are $\gamma\simeq 0.25\Omega,\,0.6\Omega,\,0.8\Omega$ for
$Q=0.18,\,0.14,\,0.12$, respectively. GI is stabilized by magnetic
pressure for sufficiently small $\beta$. The critical
field strength for stabilizing GI increases with increasing
self-gravity, consistent with \cite{nakamura83}.   
For $Q=0.18$, GI is stabilized for $\beta \lesssim 15$. Nevertheless, 
the MRI branch for $\beta < 15$ becomes self-gravitating, so that
density perturbations still grow, even though GI does not
formally operate.  

%critical beta increases with $Q$

The GI and MRI branches only interact when their growth rates
are similar. This is seen in Fig. \ref{compare_growth3} for $Q=~0.18$
where the GI branch approaches a MRI branch at $\beta\simeq 25,\,
\gamma\simeq 0.2\Omega$. In fact, following the red curve to smaller
$\beta$ indicates GI transitions to MRI. The
`gaps' in the GI and MRI branches for $Q=0.18$ and $Q=0.12$ may be due
to the phenomenon of avoided crossing, as seen in stars
\citep[e.g.][]{aizenman77} and accretion tori/disks
\citep[e.g.][]{christo93,ogilvie98}, where physically distinct modes approach
one another in frequency and exchange character. However, we cannot
exclude the possibility  that some modes may have been missed in a 
numerical search of eigenfrequencies.  
%maybe ogilvie?

Thus, our results do not rigorously prove that the GI and MRI branches
do not intersect.  Nevertheless, the continuous variation of $\tau$
strongly suggest that unstable modes can transition smoothly from 
MRI to GI and vice versa, especially at low $\beta$.     
%no unique transition

\subsubsection{Case study} 
\begin{figure}
  \includegraphics[width=\linewidth]{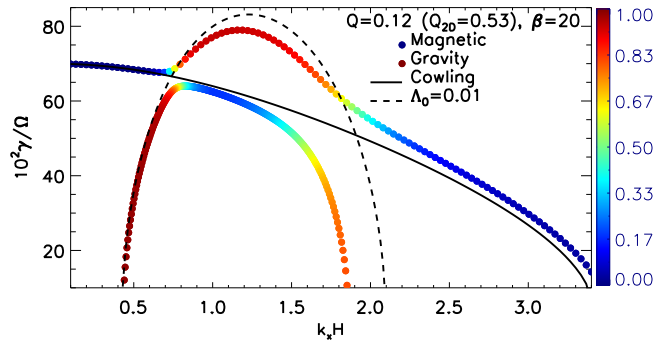}  
  \caption{Growth rates of unstable modes in the massive isothermal
    disk with $Q=0.12$ and $\beta=20$, as a function of the horizontal
    wavenumber $k_x$. The colorbar measures the importance of
    self-gravity by $\tau$. The solid line corresponds to MRI modes in the
    Cowling approximation. The dashed line corresponds to pure GI
    modes, obtained by including a high resistivity in the full
    problem. 
    \label{compare_growth3_Q01d2}}
\end{figure}

In reality, perturbations with a range of $k_x$ will be present for a
given set of disk parameters. Fig. \ref{compare_growth3_Q01d2} show
growth rates as a function of $k_x$ in a disk with
$Q=0.12,\,\beta=20$, where MRI and GI have comparable growth
rates. All perturbations with $k_xH \lesssim 3.5$ grow dynamically
($\gamma\gtrsim 0.1\Omega$, or $\lesssim$ 1.6 orbits).

We also plot in Fig. \ref{compare_growth3_Q01d2} growth rates obtained 
from the Cowling approximation, which isolates MRI; and that from a
high-resistivity run, which isolates GI by allowing the field lines to
slip through the fluid. We refer to these as  
pure MRI and pure GI, respectively.  
For $k_xH\lesssim 0.7$, growth rates are equal to
those on the pure MRI and pure GI branches. That is, MRI and GI
operate independently until their growth rates become equal as a
function of $k_x$.  

The dispersion relation $\gamma(k_x)$ deviates from the
pure GI/MRI curves with increasing $k_x$, implying stronger interaction between
magnetic and density perturbations. Comparing pure GI (dashed line)
and the gravitationally-dominated portions of $\gamma(k_x)$ shows that 
inclusion of magnetic field stabilizes high-$k_x$ pure GI. (Note also
the slight decrease in the most unstable $k_x$.) 
This stabilization is due to magnetic pressure \citep{lizano10},  
consistent with pressure stabilizing small-wavelength GI only.   

Comparing pure MRI (solid line) and the magnetically-dominated
portions of $\gamma(k_x)$ show that self-gravity increases MRI growth
rates at large $k_x$. This effect is small but noticeable, which can
be used as a code test for non-linear simulations. Note that this
destabilization by self-gravity is through the linear response, rather 
than through the background stratification (which is stabilizing).  

%smooth transitions, mode exchange

\section{Effect of an azimuthal field}\label{result3}
In this section we use the setup of case 2 described in
\S\ref{linear}, and examine the effect of an azimuthal field so that
$B_y\neq 0$,   
parametrized by $\epsilon \equiv B_y/B_z$. However, we continue to
use $B_z$ for normalizations and $\beta$ is associated with the
vertical Alfven speed. We also extend the previous calculations to the
full disk $z\in[-Z_s,Z_s]$,  which allows us to compare the effect of
self-gravity on MRI modes with different symmetries across the
midplane. We use an isothermal disk throughout. 

\subsection{Ideal disks with MRI} 
We consider disks with $Q=0.2$ ($Q_\mathrm{2D}=~0.72$) and
$\beta=10$ in the limit of ideal MHD ($\Lambda_0=100$). Gravitational
instability is not expected because Fig. \ref{compare_growth3} shows
that even for $Q=0.18$, GI is suppressed for $\beta \lesssim 15$. 

\begin{figure}
  \includegraphics[width=\linewidth,clip=true,trim=0cm 2cm 0cm
    0cm]{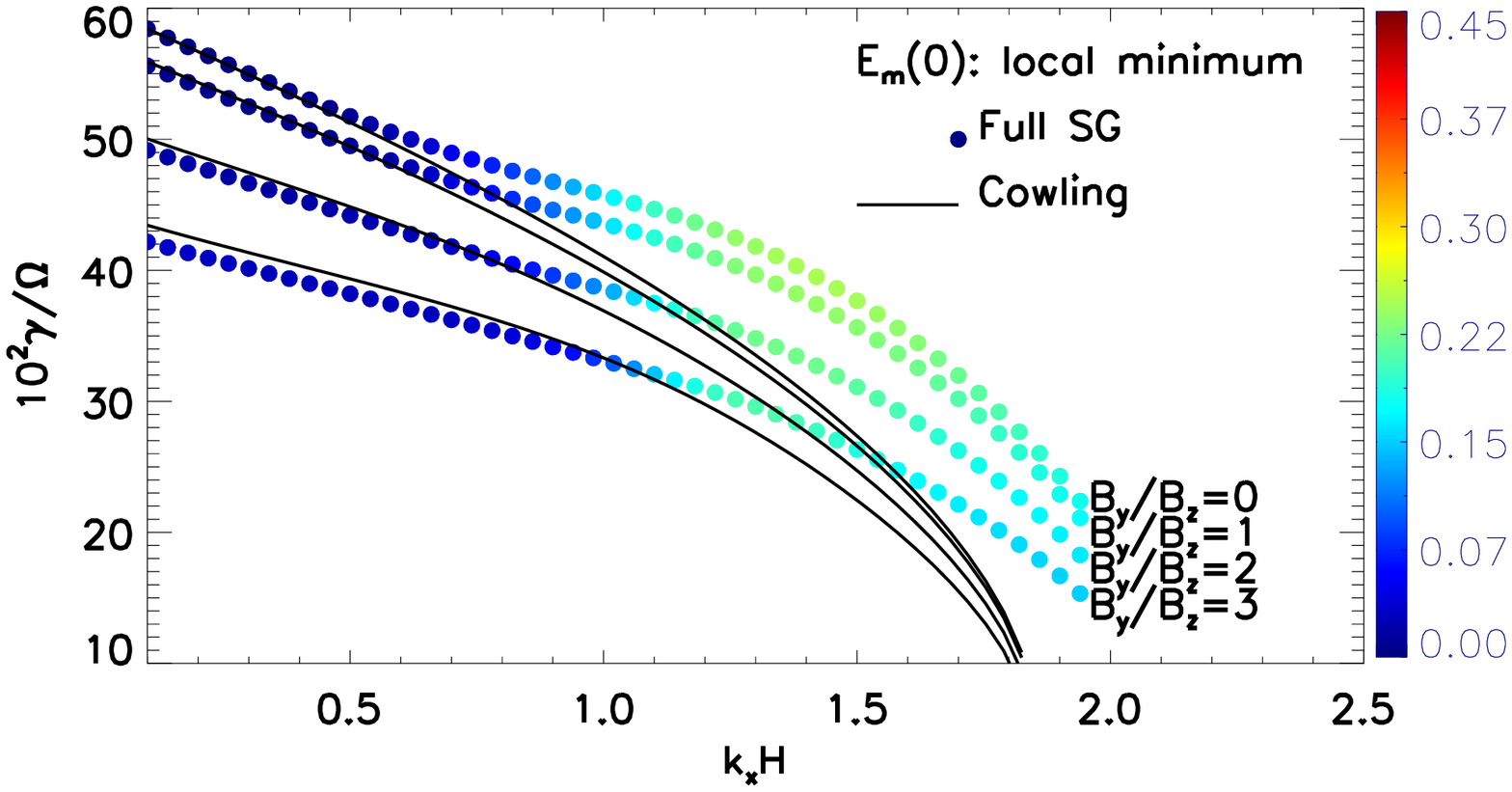}  
  \includegraphics[width=\linewidth,clip=true,trim=0cm 0cm 0cm
    0.52cm]{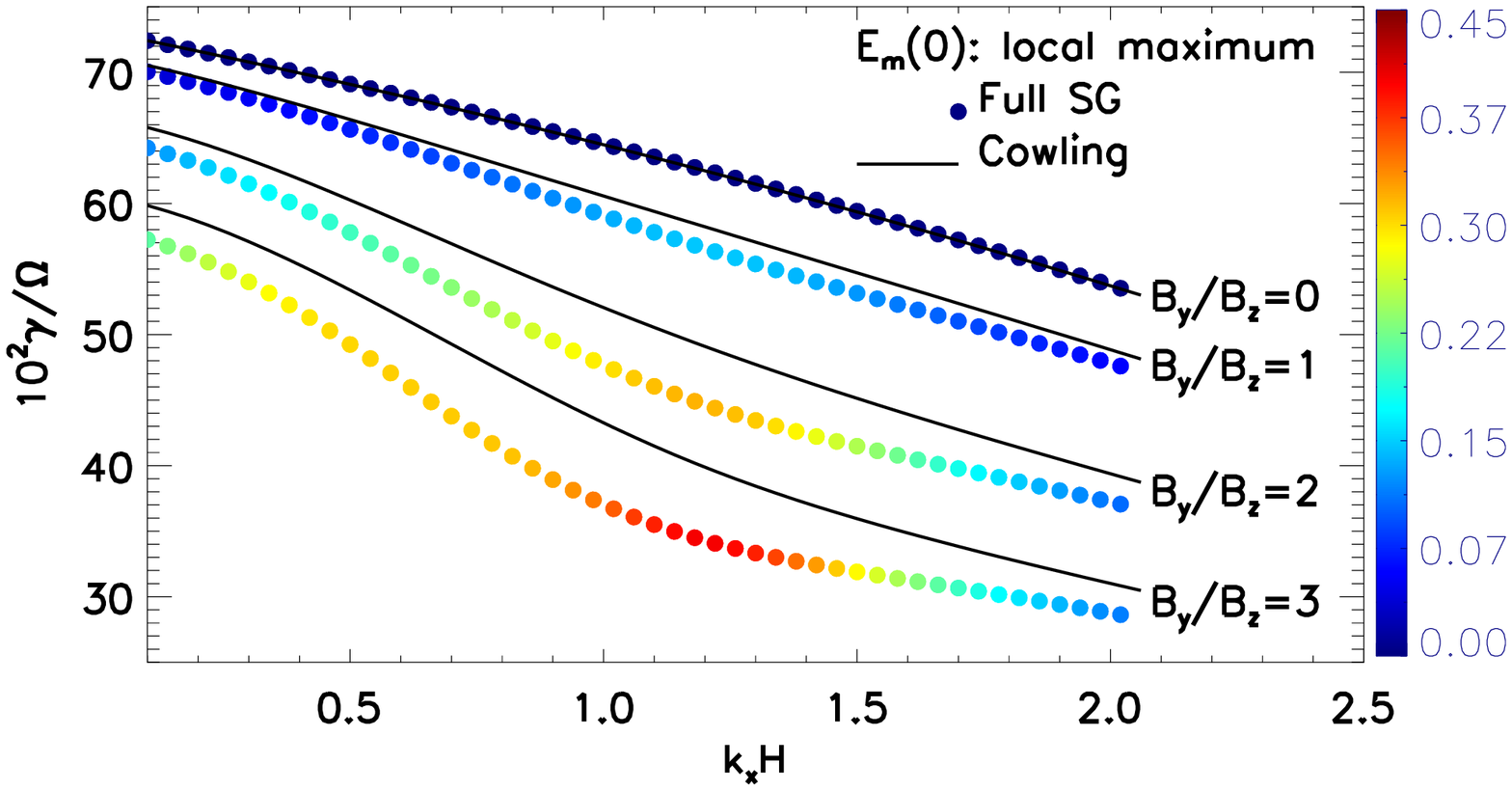} 
  \caption{MRI growth rates in isothermal disks with $Q=~0.2$ ($Q_\mathrm{2D}=0.72$) and 
    $\beta=10$ for a range of azimuthal field strengths $B_y/B_z$. The
    dots are solutions computed from the full problem, with the
    colorbar measuring the gravitational potential perturbation 
      via $\tau$, while
    the solid curves are computed from the Cowling approximation. 
    For $B_y=0$, modes in top and bottom panels have
    $W^\prime(0)=0$ and $W(0)=0$, respectively.
    \label{compare_growth3_tilted}}
\end{figure}

Fig. \ref{compare_growth3_tilted} show MRI growth rates for
$B_y/B_z=0,\,1,\,2$ and $3$. We divide the modes into two categories
depending on the extremum of magnetic energy at the midplane. 
The top panel are modes where $E_m$ has a local minimum at $z=0$ and 
the bottom panel are modes where $E_m$ has a local maximum at $z=0$. 
The latter set of modes were excluded in the previous sections by
midplane boundary conditions. 
We also plot growth rates computed in the Cowling
approximation. As expected,  $\avg{E_g}<\avg{E_m}$, so none of the
modes are energetically dominated by self-gravity.

Consider first modes in the top panel of 
Fig. \ref{compare_growth3_tilted}. As with previous results,  
self-gravity destabilizes modes with  $k_xH\gtrsim O(1)$. Consequently, the
cut-off wavenumber is larger when SG is included. 
Destabilization is most effective for purely vertical fields: with
$\epsilon=0,\, k_xH\simeq 1.4$, SG increases the growth rate by $\sim
30\%$. For $B_y=0$ we find the density perturbation $W(z)$ is an even
function. Although these modes are fundamentally magnetic, this is consistent with
\cite{goldreich65a}, who showed that SG can only destabilize 
symmetric density perturbations.  
With increasing $B_y$, we find $W$ deviates from an even
function. 
Together with the increased total magnetic pressure with 
$B_y$ (since $B_z$ is fixed), destabilization by SG weakens. 
Thus, the Cowling approximation becomes increasingly good with stronger
$B_y$ for these modes.

\begin{figure}
  \includegraphics[width=\linewidth,clip=true,trim=0cm 1.5cm 0cm
    0cm]{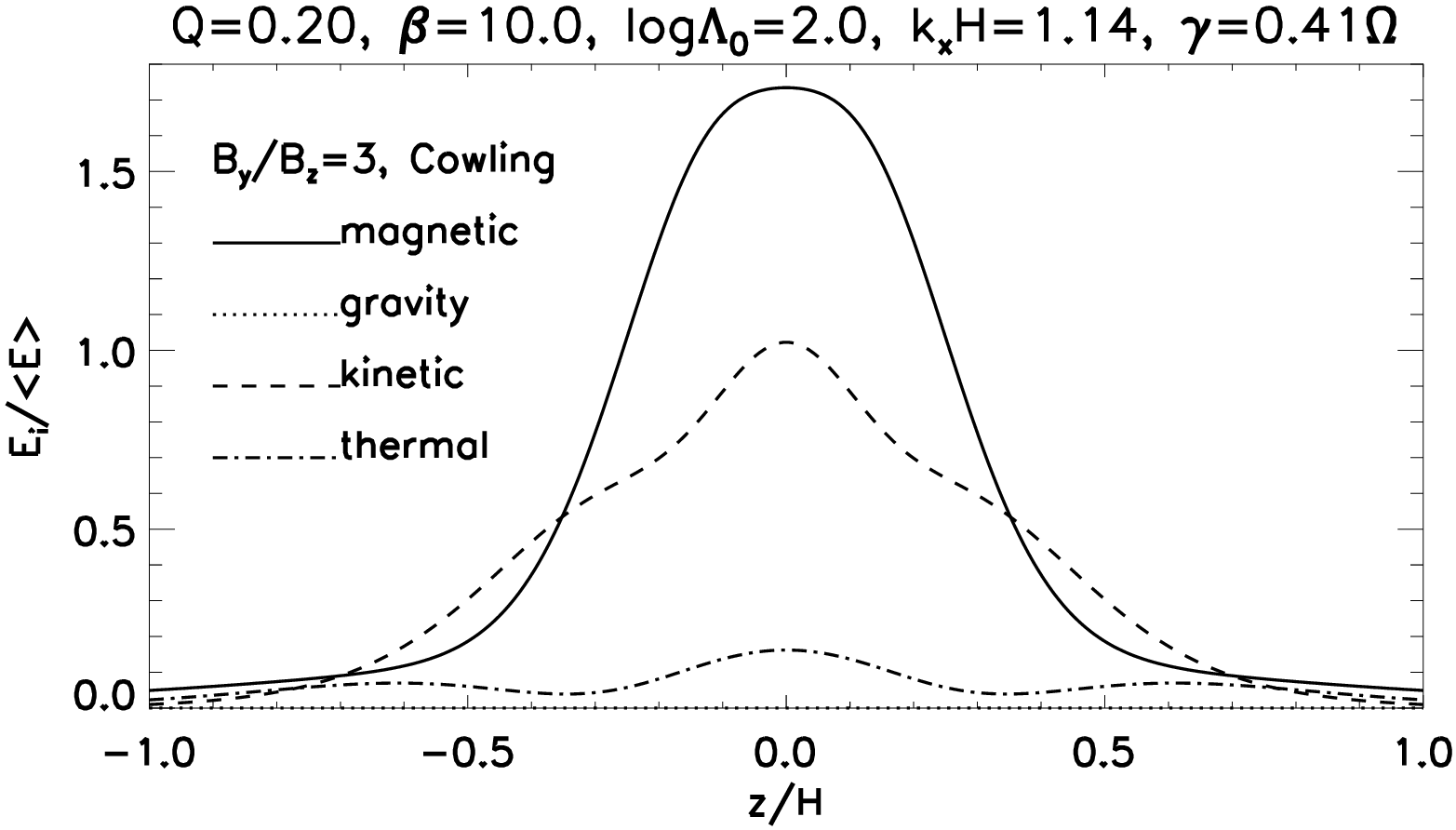}  
  \includegraphics[width=\linewidth,clip=true,trim=0cm 0cm 0cm
    0.cm]{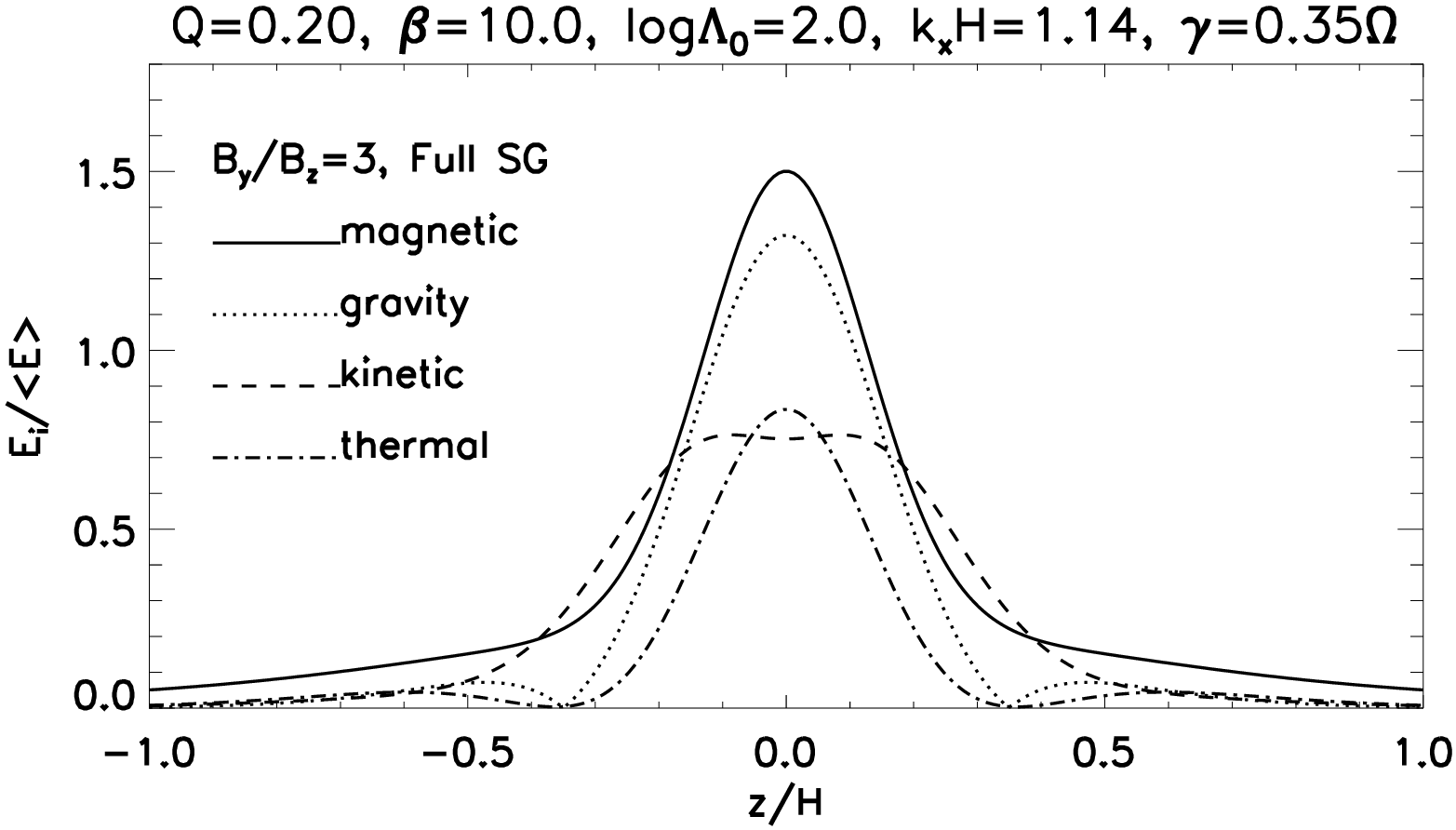} 
  \caption{Energy densities for a MRI mode in an isothermal ideal disk
    with an azimuthal field $B_y = 3B_z$, computed in the Cowling
    approximation (top) and with full self-gravity (bottom). These
    modes correspond to those in the bottom panel of
    Fig. \ref{compare_growth3_tilted}.    
    \label{result_tilted}}
\end{figure}

The modes in the bottom panel of Fig. \ref{compare_growth3_tilted} 
display opposite behavior. For $B_y=0$ we find $W(z)$ is odd, and
self-gravity has no effect. When $B_y>0$, $W$ deviates from an
odd function and the midplane density perturbation $|W(0)|$
increases. SG is stabilizing for these modes at all wavelengths, and
is most effective at $k_xH =  O(1)$. 
Fig. \ref{result_tilted} show eigenfunctions for $\epsilon=3$ and
$k_xH=1.1$ with and without the Cowling approximation.
SG significantly enhances the midplane density perturbation, making the
gravitational potential energy comparable to the magnetic energy, 
which becomes more confined near the midplane.  

To interpret the above result for modes with magnetic energy
concentrated at the midplane, we note that compressibility affects the
MRI in the presence of an azimuthal field even in a
non-self-gravitating disk. If the perturbed disk remains in vertical
hydrostatic equilibrium, then  
\begin{align} 
  |W|\sim \frac{B_y}{\mu_0\rho}|\dby|,
\end{align}
to order of magnitude in a non-SG disk. Thus a strong azimuthal field can 
cause a large density perturbation \citep{pessah05}. We checked that
for the modes in Fig. \ref{result_tilted}, vertical velocities are 
small, $|\dvz|/\left(|\dvx|^2+|\dvy|^2\right)^{1/2} \lesssim 0.2$. 

Compressibility is enhanced by an azimuthal field, which is 
stabilizing for the MRI \citep{kim00}. %because...?
This effect is significant for $\epsilon=3$ because the azimuthal
Alfven speed is sonic.      
%maybe blaes 
Fig. \ref{result_tilted} indicates that self-gravity further enhances
compressibility, and therefore stabilization. We suspect this is
overwhelmed by the destabilization effect of SG, because the density
perturbation has an anti-symmetric component.      

\subsection{Resistive disks with GI}
\begin{figure}
  \includegraphics[width=\linewidth,clip=true,trim=0cm 2cm 0cm
    0cm]{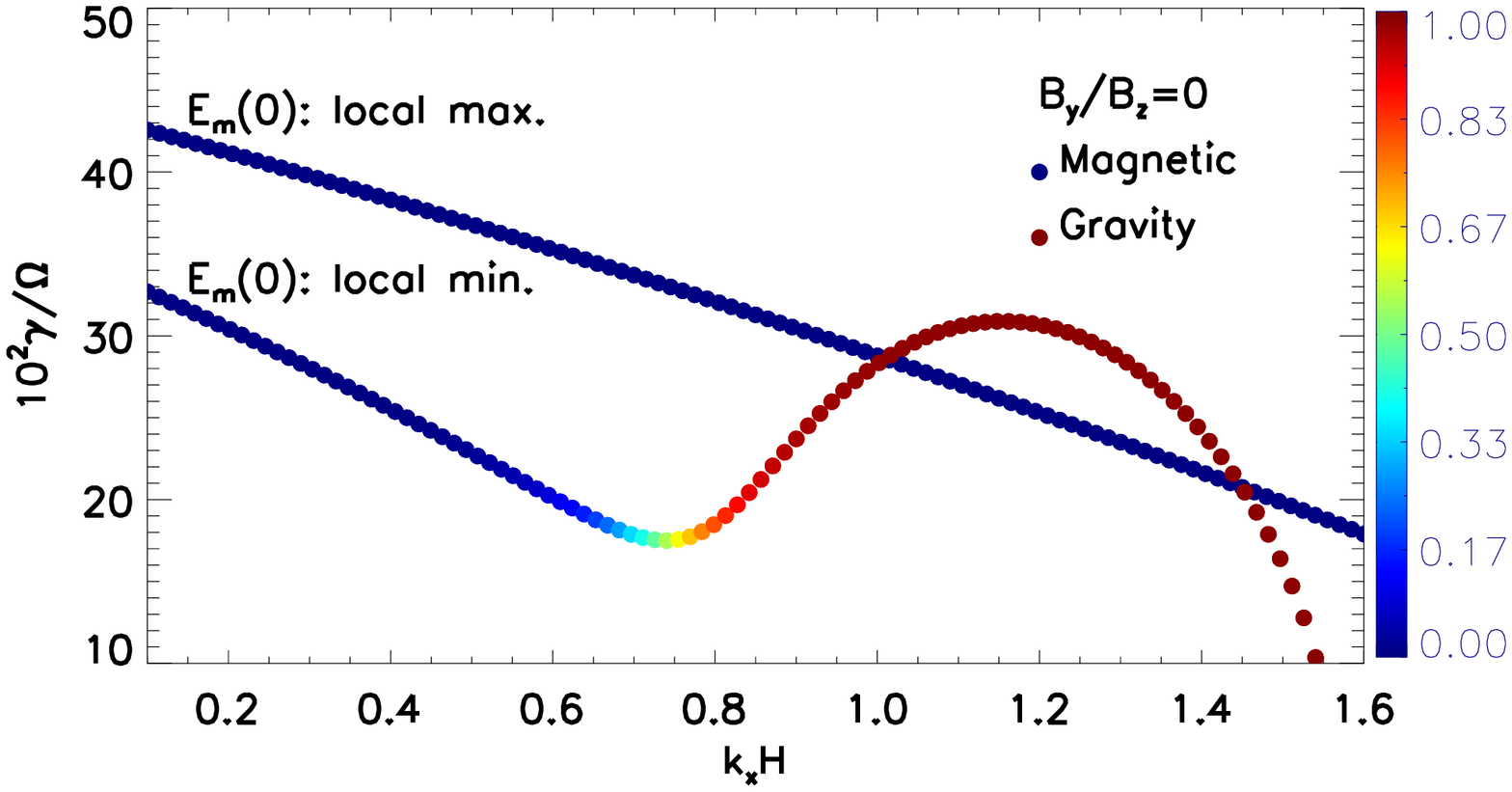}  
  \includegraphics[width=\linewidth,clip=true,trim=0cm 2cm 0cm
    0.52cm]{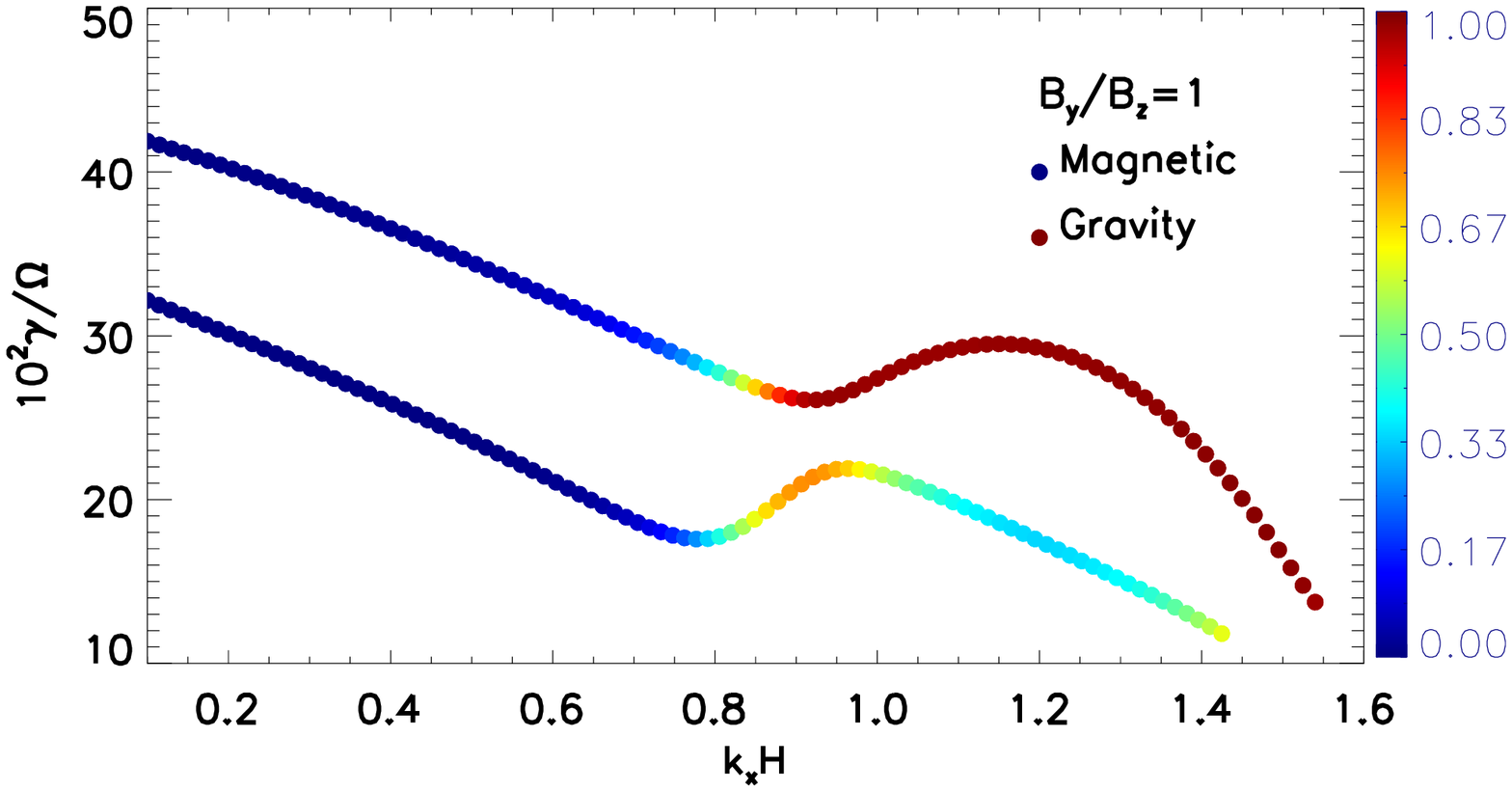}
  \includegraphics[width=\linewidth,clip=true,trim=0cm 0cm 0cm
    0.52cm]{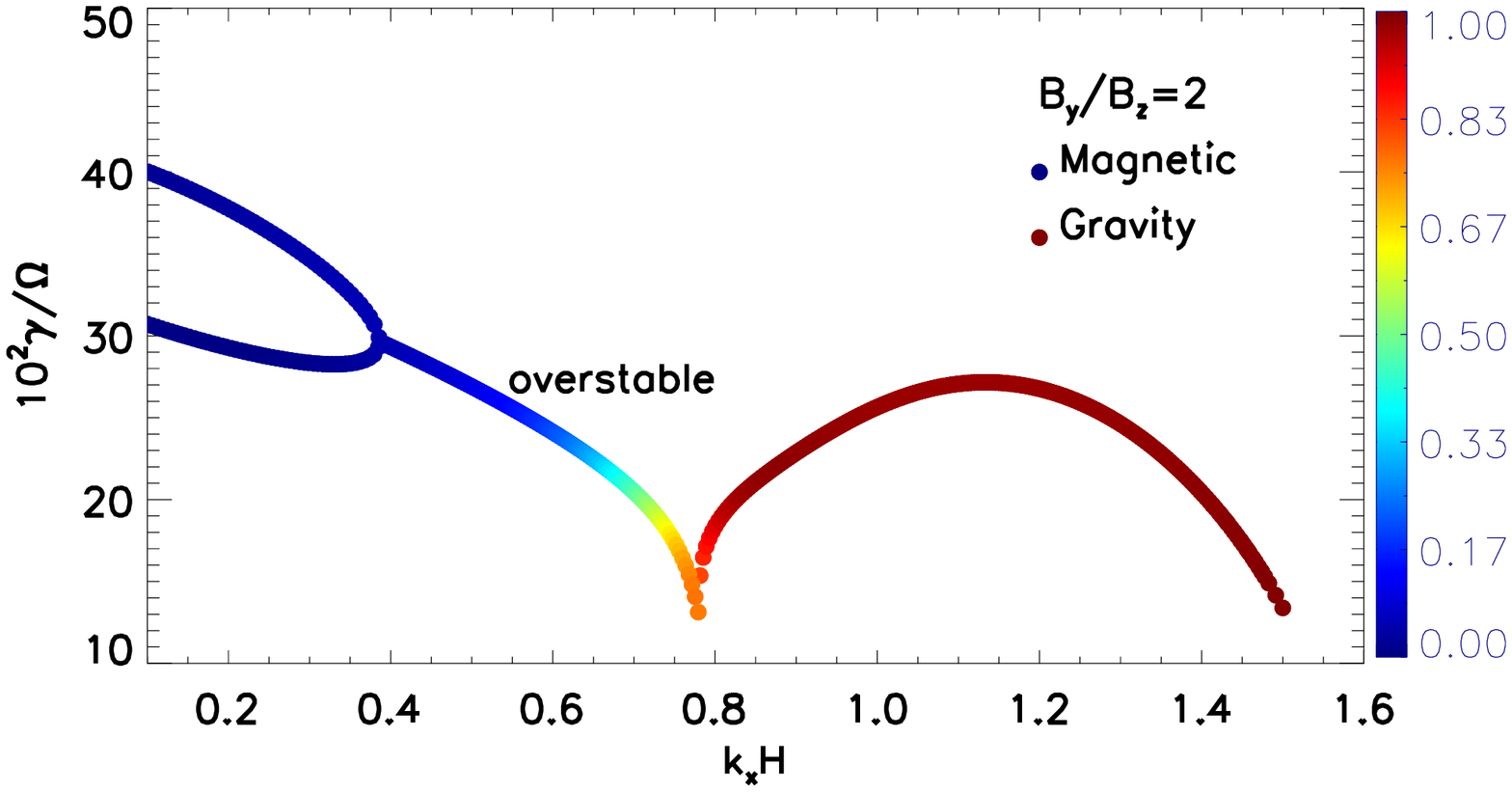}
  \caption{Growth rates in isothermal resistive disks with $Q=0.18$
    ($Q_\mathrm{2D}=0.67$), $\beta=100$ and $\Lambda_0=0.1$. For
    $B_y=0$, the upper and lower MRI modes have anti-symmetric and
    symmetric density perturbations, corresponding to $W(0)=0$ and
    $W^\prime(0)=0$, respectively.  
    %anti-symmetric density
    %    perturbation $W(0) = 0$ and symmetric density perturbations,
    %    $W^\prime(0) = 0$.  
    For $B_y/B_z=2$ the
    overstable modes have non-zero real frequencies. 
    \label{compare_growth3_tilted_resis}}
\end{figure}

Here we examine a resistive disk which permits MRI and GI by setting $Q=0.18$,
$\Lambda_0=0.1$ and $\beta=100$. 
Fig. \ref{compare_growth3_tilted_resis} show growth rates for
$\epsilon=0,\,1$ and $2$. For $B_y=0$, MRI and GI are decoupled except 
for a narrow range of $k_x$ in which the lower MRI modes transitions to
GI. Notice that the upper MRI modes intersect the GI branch. There is no
interaction because the upper MRI modes have anti-symmetric $W(z)$ whereas
the GI modes have symmetric $W(z)$. 

Introducing $B_y = B_z$ leads to an exchange in the mode
characters. For $k_xH\lesssim 0.9$ the modes on the two MRI branches
are similar to the vertical field case. However, for $k_xH\gtrsim0.9$ 
the upper MRI mode transitions to GI, for which $E_m(0)$ is a
minimum; and the lower MRI mode has $E_m(0)$ being a maximum. We find
all perturbations with  $k_xH\gtrsim0.9$ have symmetric $W(z)$. 
%small scale perts are all even 

\begin{figure}
  \includegraphics[width=\linewidth,clip=true,trim=0cm 0cm 0cm
    0cm]{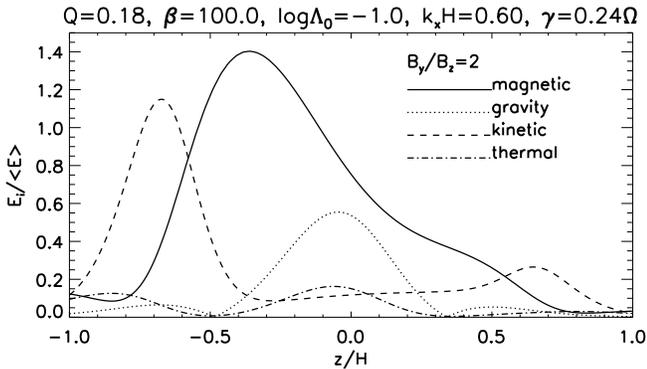}
  \caption{Overstable MRI mode in an isothermal resistive disk with
    $Q=0.18$ ($Q_\mathrm{2D} = 0.67$), $\Lambda_0=0.1$ and $\beta=100$. 
    The mode has a real frequency $\omega = 0.059\Omega$, or
    $\omega/\gamma \simeq 0.2 $.   
    \label{result_tilted_overstable}}
\end{figure}

Increasing the azimuthal field further to $B_y=2B_z$ we find
overstable MRI modes  with non-negligible real
frequencies \citep{gammie96}. An example is shown in
Fig. \ref{result_tilted_overstable}. Notice the density/potential
perturbation is off-set from the midplane. This is not possible for
pure GI \citep{goldreich65a}. Thus, these overstable MRI modes indeed
become self-gravitating, before being stabilized. 
%these may lead to significant vertical motions 

Notice also in Fig. \ref{compare_growth3_tilted_resis} the disappearance of
magnetic modes between $0.8\lesssim k_xH\lesssim 1.5$ as $B_y$ is
increased. For $B_y=2B_z$, MRI and GI are again independent 
because they operate at distinct radial scales. This implies that
perturbations unstable to GI cannot develop MRI.

\section{Summary and discussion}\label{summary}
In this paper, we have performed axisymmetric linear stability
calculations of magnetized, self-gravitating, vertically stratified
disks in the local approximation. Our models include resistivity and
azimuthal fields. We have identified regimes under which the
magneto-rotational instability (MRI) is affected by disk self-gravity
(SG).

For a vertical field, the requirement for the MRI to operate is that   
its vertical wavelength $\lambda \lesssim 2H$.  
%While $\lambda$ is independent of
%the strength of self-gravity $Q$ {\bf for fixed $\beta$}, 
The disk thickness $H=H(Q)$ decreases with increasing SG. This 
reduces MRI growth rates when 
$\beta$, and hence $\lambda$, is fixed.  
Thus, a sufficiently massive disk can potentially suppress the MRI. 
The MRI is also restricted to larger radial 
scales as $Q$ is lowered. This means that the MRI becomes more global
in self-gravitating disks. 

  The condition $\lambda < 2H$ 
  %requires $f(Q)/\sqrt{\beta}$ to
  %  be sufficiently small (see
  %  Eq. \ref{lambda_ideal}---\ref{lambda_resis}), so larger values of
  %  $\beta$ are required to  
  %  permit the MRI with increasing self-gravity, since $f(Q)$ increases
  %  with decreasing $Q$. 
  may be written more precisely as 
  \begin{align}\label{cond1}
    \frac{n}{\mathrm{min}(\Lambda_0, 1)}\frac{f(Q)}{\sqrt{\beta}} \lesssim 1,
  \end{align}
  where $n\sim 3$ is a numerical factor and
  $\mathrm{min}(\Lambda_0,1)$ accounts for the ideal and resistive
  limits (see Eq. \ref{lambda_ideal}---\ref{lambda_resis}). Since $f$
  increases with decreasing $Q$, Eq. \ref{cond1} implies
  the MRI requires larger values of $\beta$ with  
  increasing self-gravity. 
  For definiteness, consider the ideal polytropic disk. Then Eq. \ref{cond1} is
  \begin{align}\label{cond2}
    \beta^{-1/2} \lesssim
    \frac{\sqrt{15}}{4\pi}\sqrt{Q}\arccos{\left(\frac{Q}{1+Q}\right)}. 
  \end{align}
  For a non-self-gravitating disk, $Q\to\infty$ and Eq. \ref{cond2} is
  $\beta \gtrsim 16\pi^2/30\simeq 5$. For $Q\ll 1$, the condition is
  $\beta \gtrsim 64/15Q$, giving $\beta\gtrsim 20$ for
  $Q=0.2$. We confirm this numerically, finding the MRI growth rate
  $\gamma  \lesssim 0.1$ when $\beta \lesssim 3.3$ for $Q=20$ and 
  $\beta \lesssim 17$ for $Q=0.2$.  
%  Inserting $Q\sim 0.2$ for an ideal disk marginally stable to
%  self-gravity, we find $v_{A0}\lesssim 0.2\csmid$ for the MRI. 

  We can also place an upper bound on the absolute field strength
  $B_z$. Writing $v_{A0} = B_z\sqrt{4\pi G Q/\mu_0\Omega^2}$, we find 
  Eq. \ref{cond2} is independent of $Q$ for $Q\ll1$, and 
  \begin{align}\label{cond3}
    \frac{B_z}{\csmid\Omega}\sqrt{\frac{\pi G}{\mu_0}}\lesssim
    \frac{\sqrt{15}}{16}
  \end{align}
  is needed for the MRI to operate in the ideal polytropic disk with strong
  self-gravity. Although both the MRI wavelength and disk thickness
  vanish as $Q\to 0$, the MRI can still operate provided the field is
  sufficiently weak according to Eq. \ref{cond3}.  

%Q->0, thickness -> 0 

%This condition translates to an upper
%  limit on the actual field strength $B_z$. 
%
%  For definiteness, consider
%  the polytropic disk with $Q \ll 1$. The disk thickness is then 
%  $H\sim \csmid\pi\sqrt{Q}/2\Omega$. This applies at marginal
%  gravitatational stability for which $Q\sim 0.2$. Next, expressing
%  $\rho_0$ in terms of $Q$, we have $\lambda\sim 4\pi B_z \sqrt{\pi
%    GQ/\mu_0}/\Omega^2$. Then 
%  \begin{align}
%    \frac{B_z}{\csmid\Omega}\sqrt{\frac{\pi G}{\mu_0}}\lesssim
%    \frac{1}{4}
%  \end{align}
%  is needed for the MRI when self-gravity is strong. 
  % We can quanitfy this by comparing
  %$\lambda$ to the Jeans length $\lambda_J \equiv 2
  %c_s^2/G\Sigma\sim 2\csmid^2/G\rho_0 H \sim 8\pi f^2Q H$. Then $\lambda
  %< H $ translates to
  %\begin{align}    
  %\frac{\lambda}{\lambda_J}\lesssim \frac{1}{8\pi f^2(Q) Q}.
  %\end{align}
  %Given a mid-plane density $\rho_0$, and hence $Q$, the left-hand-side
  %is a measure of the field strength. 

Interestingly, for layered resistivity we do not find layered
magnetic perturbations when the disk is massive. This is 
consistent with the MRI becoming vertically global with increasing self-gravity.    
For non-self-gravitating disks $\lambda\ll H$, so the MRI can be
restricted to regions of size $L<H$, i.e. an active layer. This not
compatible with $\lambda \sim H$, as found for massive disks. Hence we
find magnetic perturbations penetrate into the high-resistivity dead
zone (e.g. $Q=0.2$ in Fig. \ref{poly_layer}), and there is no distinct
boundary between active and dead layers. This suggests that the
picture of layered accretion  \citep[e.g.][]{fleming03} may not be applicable 
to self-gravitating disks.  

%SG through 
We find MRI modes with radial scales of $\sim H$ can acquire  
density perturbations in massive but Toomre-stable disks. 
This occurs when the MRI is weak, for example with a strong field or
high resistivity. We argue in that case $\lambda\sim H$, so the MRI is
compressible and the associated density
perturbation can be enhanced by self-gravity.
%Our 
%interpretation involves the MRI to provide a seed density
%perturbation %in the absence of self-gravity. 

%what is required? a seed pert, plus amplification

%further enhancing density perturbations. 
At this point it is worth mentioning previous non-linear simulations 
of magnetized self-gravitating galactic and circumstellar disks
\citep{kim03,fromang04a,fromang04}. These authors find self-gravity
did not enhance MRI density fluctuations significantly. However, they
employed ideal MHD simulations with gas-to-magnetic pressure ratios of
order $10^2$ to $10^3$. %MRI is basically incompressible 
This is qualitatively consistent with our results, as 
self-gravity is not expected to influence the MRI in this  
regime of $\beta$, except through the background state. 
%both `seed' and amplfication are small 
For example, \cite{fromang04} found
MRI turbulence is more coherent in self-gravitating disks. This may be related
to our finding that small radial scale MRI is suppressed when
self-gravity is included in the background equilibria. 

Physically, we expect MRI to interact with self-gravity when
their spatial scales are similar. Because self-gravity acts globally in the
vertical direction, for it to affect the MRI, future non-linear
simulations should consider parameter regimes in which the MRI is
vertically global. %this makes MRI more compressible
Indeed, in the setup of \cite{kim03}, the disk
scale height exceeds the MRI vertical wavelength and self-gravity has
little impact.

%GI/MRI: avoided crossings, relevance to simulations
Curiously, when GI and MRI are simultaneously supported, we find
unstable modes transition between MRI and GI. There exists modes 
with comparable potential and magnetic energy perturbations, which 
demonstrates MRI and GI can interact. These 
transitions occur smoother with decreasing $\beta$
(Fig. \ref{compare_growth3}) or increasing $k_x$
(Fig. \ref{compare_growth3_Q01d2}). The latter implies that, in order
to capture the magneto-gravitational interactions represented by these
intermediate modes, non-linear simulations must resolve radial scales
smaller than the most unstable GI mode. For example, 
Fig. \ref{compare_growth3_Q01d2} suggest radial scales down to $\sim H/2$
should be well-resolved. 

%azimuthal fields 

We examined the effect of an additional azimuthal field, while
keeping the vertical field at fixed strength. In this case, we also
relaxed the equatorial symmetry condition applied previously and
considered the full disk column. Self-gravity affects the MRI
differently depending on its character. 
%midplane symmetry of the MRI 
%density
%perturbation. 
Self-gravity destabilizes MRI modes where the magnetic
energy has a minimum at $z=0$, these modes have a symmetric
density perturbation in the limit $B_y\to0$. However, 
self-gravity stabilizes MRI modes where the magnetic energy has a
maximum at $z=0$, these modes have an anti-symmetric density
perturbation in the limit $B_y\to0$. This stabilization effect is
stronger for increasing $B_y$. Previous linear calculations show that 
increased compressibility associated with a toroidal field stabilizes
the MRI \citep{kim03}. We conjecture that self-gravity 
further enhances this effect. Non-linear MRI simulations with strong
toroidal fields that neglect self-gravity may over-estimate the
strength of MRI turbulence.   
%can't do WKB (can't tell symmetries)

\subsection{Caveats and outlooks}
We discuss below two major extensions to our linear model that
should be undertaken, before embarking on non-linear simulations of
magnetized, self-gravitating disks, which is our eventual goal.  

\emph{Beyond the shearing box.} The shearing box ignores the curvature
of toroidal field lines present in the global disk 
geometry. \cite{pessah05} demonstrated new effects on the MRI 
when the curvature of a super-thermal toroidal field is accounted for;
although \citeauthor{pessah05} focused on modes with large (small)
vertical (radial) wavenumbers, for which we expect self-gravity
can be ignored. 
Since compressibility becomes important for strong
toroidal fields, the effect of self-gravity on modes with $k_xH\sim1$
may become significant when super-thermal toroidal fields are
considered. However, global disk models will be necessary
to self-consistently probe this regime.

\emph{Beyond axisymmetry.} Axisymmetric perturbations, as we have
assumed, preclude gravitational torques \citep{lynden-bell72}. 
The local non-axisymmetric stability of magnetized self-gravitating
thin disks has been considered by several authors
\citep{elmegreen87,gammie96b,fan97,kim01}. However, two-dimensional models
exclude the MRI. It will be necessary to generalize these studies to
3D in order to investigate the impact of the MRI on angular momentum
transport by gravitational instability. Furthermore, self-gravitating
disks can develop global spiral instabilities while stable against local 
axisymmetric perturbations \citep{papaloizou89,papaloizou91}. Global
non-axisymmetric linear models will be desirable to support non-linear
simulations of this kind \citep{fromang04c,fromang05}.

\acknowledgements
I thank K. Menou, S. Fromang and A. Youdin for helpful discussions
during the course of this project. The project source codes may be found at
\url{https://github.com/minkailin/sgmri}. 

\appendix
\section{Analytic equilibrium for the polytropic disk}\label{appen1}
\begin{figure}
  \includegraphics[width=\linewidth]{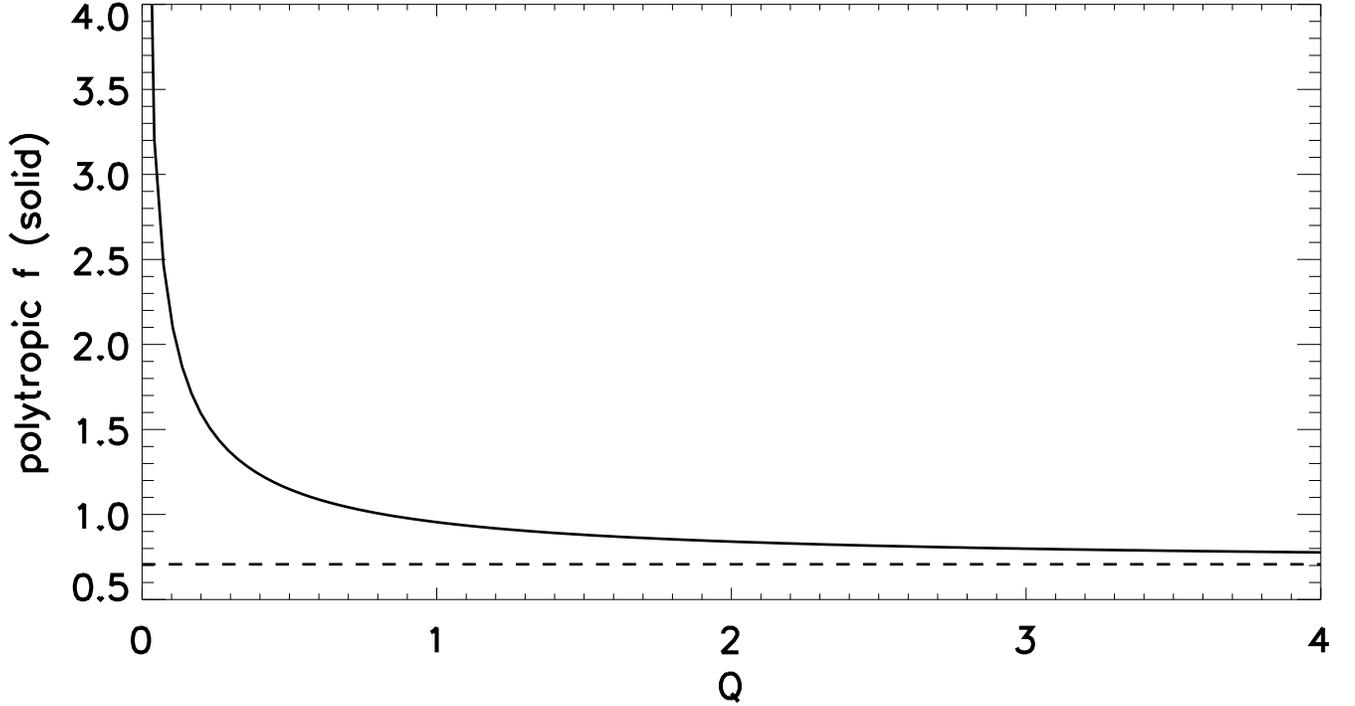}
  \caption{The function $f(Q)$ describing vertical hydrostatic 
    equilibrium in self-gravitating polytropic disks (solid line). The
    horizontal dashed line is the asymptotic value of $1/\sqrt{2}$ for
    large $Q$.  
    \label{plot_fq}}
\end{figure}

For a polytropic disk with $P = K\rho^2$ the dimensional equilibrium equation to
be solved is 
\begin{align}
  0=\csmid^2\frac{d^2}{dz^2}\left(\frac{\rho}{\rho_0}\right) +
  \Omega_z^2 + \frac{\Omega^2}{Q}\left(\frac{\rho}{\rho_0}\right), 
\end{align}
which is obtained by combining Eq. \ref{eqm_eqns1} and
\ref{eqm_eqns2} with the above equation of state. The solution is
\begin{align}  
  \frac{\rho}{\rho_0} = \left(1 + \frac{\Omega_z^2}{\Omega^2}Q\right)\cos{\left(a z\right)} - 
  \frac{\Omega_z^2}{\Omega^2}Q,  
\end{align}
where  
\begin{align}
  a^2 \equiv \frac{\Omega^2}{Q\csmid^2}. 
\end{align}
The polytropic disk thickness is
\begin{align}
  H =
  \frac{\csmid}{\Omega}\sqrt{Q}\arccos\left(\frac{\Omega_z^2Q}{\Omega^2+ \Omega_z^2Q}\right).  
\end{align}
Given a fixed mid-plane temperature, the function $f(Q)\equiv
\csmid/\Omega H$ is an inverse measure of the disk thickness, and $f$ increases with decreasing $Q$,
as shown in Fig. \ref{plot_fq}. This corresponds to a thinner disk
with increasing strength of vertical self-gravity.

\section{Relation between $Q$ and the Toomre parameter}\label{q3d2d}
\begin{figure}
  \includegraphics[width=\linewidth]{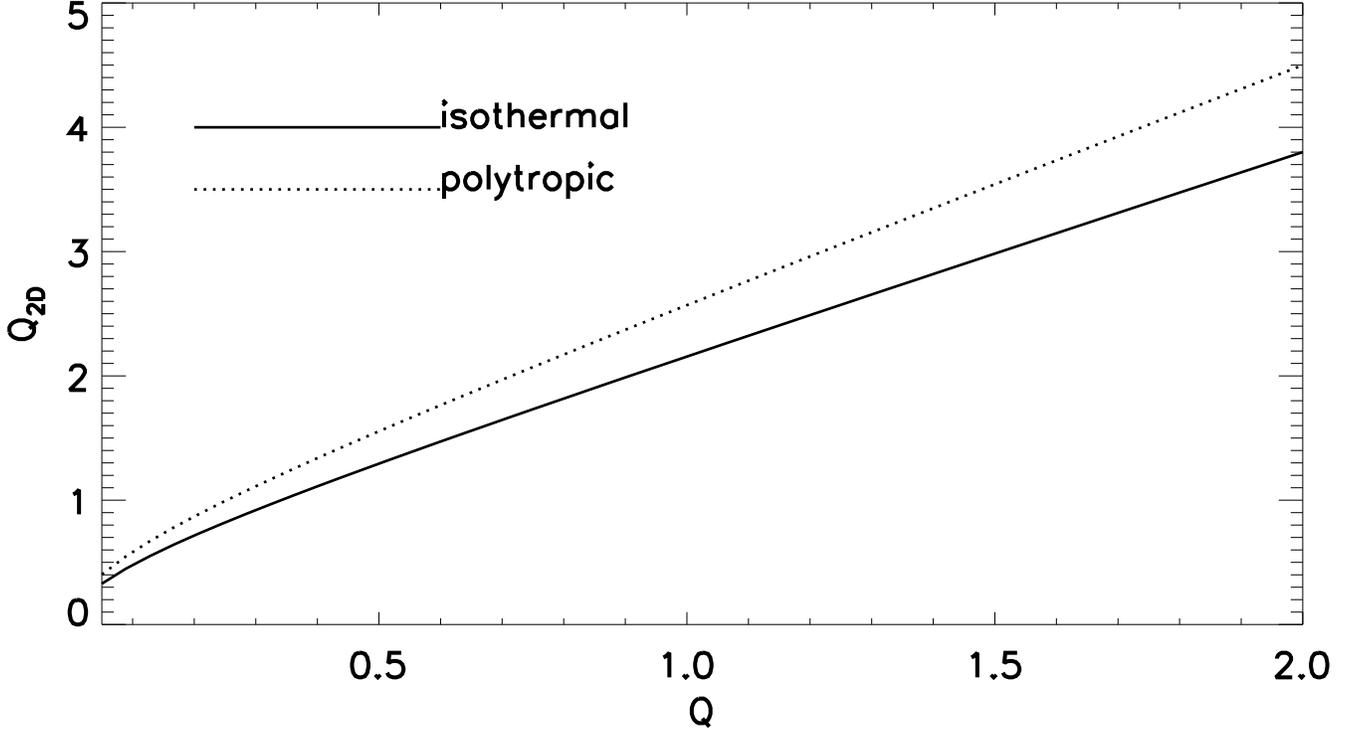}
  \caption{Relation between the self-gravity parameter $Q$ used in
    this paper and the Toomre parameter $Q_\mathrm{2D}$ for razor-thin
    disks. 
    \label{plot_q3d2d}}
\end{figure}

The Toomre parameter defined for razor-thin disks is 
\begin{align} 
  Q_\mathrm{2D}\equiv \frac{\kappa c_s}{\pi G\Sigma},  
\end{align}
where $\Sigma$ is the total column density. To relate our
self-gravity parameter $Q$ and $Q_\mathrm{2D}$, we replace $c_s$ by
$\overline{c_s}\equiv\int\rho c_s dz/\int\rho dz$, and 
$\kappa$ by $\Omega$, giving
\begin{align}
  Q_\mathrm{2D} = 2Qf \frac{\int_0^1
    \hat{\rho} \hat{c}_s
    d\hat{z}}{\left(\int_0^1 \hat{\rho} d\hat{z}\right)^2},   
\end{align}
where each term on the right-hand-side is non-dimensionalized (see
\S\ref{non-dim}). Fig. \ref{plot_q3d2d} plots this relation 
for isothermal and polytropic disks.  

\section{Reduction to linear hydrodynamics}\label{reduction} 
Our task here is to remove the magnetic field and vertical velocity
perturbations from the linearized equations. Let us first define operators
\begin{align}
  D_0 = 1, \quad D_1 = \frac{\rho^\prime}{\rho} + \frac{d}{dz}, \quad
  D_2 = \frac{\rho^{\prime\prime}}{\rho} +
  \frac{2\rho^\prime}{\rho}\frac{d}{dz} + \frac{d^2}{d z^2},
\end{align}
and 
\begin{align}
  \overline{D}_0 = \eta D_0, \quad \overline{D}_1 = \eta^\prime D_0 + \eta
  D_1,\quad \overline{D}_2 = \eta^{\prime\prime} D_0 + 2\eta^\prime D_1 +
  \eta D_2. 
\end{align}
And we define the variables
\begin{align}
  &U \equiv \imgi\sigma\dvx - 2\Omega\dvy + \imgi k_x \w,\\
  &V \equiv \imgi\sigma\dvy + \frac{\kappa^2}{2\Omega}\dvx.
\end{align}

We first express the continuity equation in terms of horizontal
velocity, density and potential perturbations. 
The vertical velocity perturbation is 
\begin{align}
  \dvz = \frac{\imgi}{\sigma}\left(\w^\prime + \epsilon V\right),  
\end{align}
where the linearized $y$ momentum equation was used (i.e. eliminating
$\dby^\prime$ between Eq. \ref{lin_vy} and Eq. \ref{lin_vz}). 
Inserting this into the linearized continuity equation
(Eq. \ref{lin_cont}), we obtain
\begin{align}
0 = W^{\prime\prime} + \left(\ln{\rho}\right)^\prime W^\prime +
\frac{\sigma^2}{c_s^2}W +  \dphi^{\prime\prime} + \left(\ln{\rho}\right)^\prime \dphi^\prime
 + \sigma k_x \dvx + \epsilon D_1 V.
%  0=  W^{\prime\prime} + \left(\ln{\rho}\right)^\prime W^\prime +
%  \frac{1}{c_s^2f^2}\left(\frac{\rho}{Q} + \sigma^2\right)W +
% \left(\ln{\rho}\right)^\prime\dphi^\prime + k_x^2\dphi + \frac{\sigma
%    k_x}{f}\dvx.\label{final_w}
\end{align}

Next, we examine separately the cases of a vertical field with
variable resistivity and that of a tilted field with uniform
resistivity. (A similar procedure can be performed in
the general case of a tilted field with variable resistivity.) 

\subsection{Vertical field with variable resistivity}
First consider $\epsilon = 0$ and $\eta=\eta(z)$ in the linearized
equations. Denoting the $n^\mathrm{th}$ vertical derivative as $^{(n)}$, the
equations of motion give
\begin{align}
  &\dbx^{(n)} = \frac{\mu_0\rho}{B_z}D_{n-1}U + \imgi k_x \dbz^{(n-1)},\label{bx_eq}\\
  &\dby^{(n)} = \frac{\mu_0\rho}{B_z}D_{n-1}V,\label{by_eq}
\end{align}
for $n\geq1$. Differentiating the divergence-free condition for the magnetic
field gives
\begin{align}
  \imgi k_x \dbx^{\prime} + \dbz^{\prime\prime} = 0.   
\end{align}
We insert the expression for $\dbx^\prime$ from Eq. \ref{bx_eq} and the
expression for $\dbz^{\prime\prime}$ from the $z$ component of the
linearized induction equation (Eq. \ref{induct_vert}) to obtain
\begin{align}\label{bz_eq}
  -\sigma\dbz^{(n)} = k_x B_z \dvx^{(n)} + k_x\frac{\mu_0\rho}{B_z}\overline{D}_nU. 
\end{align}

%\subsection{Eliminating $\dbx$}
Inserting the above expressions for $\dbx^{\prime\prime}$,
$\dbx^\prime$ (Eq. \ref{bx_eq}) and $\dbz^\prime$  (Eq. \ref{bz_eq})
into the right-hand-side of the $x$-induction equation (Eq. \ref{induct_x} ) gives   
\begin{align}
  \imgi\sigma \dbx =
  B_z\dvx^\prime + \frac{\mu_0\rho}{B_z}\overline{D}_1U. \label{bx_expression}
\end{align}
($\bar{\sigma}\neq0$ has been assumed to obtain this.) We
differentiate this expression with respect to $z$ and eliminate the
resulting $\dbx^\prime$ using Eq. \ref{bx_eq}, to obtain
\begin{align}
%  0 = B_z\dvx^{\prime\prime} -
%  k_x^2B_z\dvx + \frac{\mu_0\rho}{B_z}\left(\overline{D}_2 - k_x^2 \overline{D}_0
%    - \imgi\sigma D_0\right)U. \label{final_vx} 
  0 = v_A^2\left(\dvx^{\prime\prime} -
  k_x^2\dvx\right) + \left(\overline{D}_2 - k_x^2 \overline{D}_0
    - \imgi\sigma D_0\right)U. \label{final_vx} 
\end{align}

%\subsection{Eliminating $\dby$}
We follow a similar procedure as above to remove $\dby$. 
We use Eq. \ref{bx_expression} and Eq. \ref{by_eq} to 
eliminate $\dbx,\dby^\prime$ and $\dby^{\prime\prime}$ from the
right-hand-side of the $y$-induction equation (Eq. \ref{induct_y}), 
%\begin{align}
%  &\imgi\bar{\sigma}\dby = f\dvy^\prime - \imgi
%  S\frac{\dbz^\prime}{k_x} +
%  \eta\dby^{\prime\prime}+\eta^\prime\dby^\prime.  
%\end{align}
%We can now insert expressions for the magnetic field derivatives using
%Eq. \ref{by_eq} and \ref{bz_eq} to obtain an expression for $\dby$, 
\begin{align}
  \imgi\bar{\sigma}\dby = B_z\dvy^\prime + \frac{\imgi S}{\sigma}\left(
    B_z\dvx^\prime + \frac{\mu_0\rho}{B_z}\overline{D}_1U\right) +
  \frac{\mu_0\rho}{B_z}\overline{D}_1V. \label{by_expression}
\end{align}
We differentiate this expression with respect to $z$, then eliminate 
$\dby$ and  $\dby^\prime$ from the left-hand-side of the resulting
expression using Eq. \ref{by_expression} and 
Eq. \ref{by_eq}, respectively. We obtain
\begin{align}
0 = v_A^2\left(\dvy^{\prime\prime} -
\frac{\sbar^\prime}{\sbar}\dvy^\prime\right)  
+ \frac{\imgi S v_A^2}{\sigma}\left(\dvx^{\prime\prime} -
\frac{\sbar^\prime}{\sbar}\dvx^\prime\right) + \frac{\imgi
  S}{\sigma}\left(\overline{D}_2 -
    \frac{\sbar^\prime}{\sbar}\overline{D}_1 \right)U 
+\left(\overline{D}_2 -
    \frac{\sbar^\prime}{\sbar}\overline{D}_1 - \imgi\sbar D_0 \right)V. \label{final_vy}
%0 = B_z\left(\dvy^{\prime\prime} -
%\frac{\sbar^\prime}{\sbar}\dvy^\prime\right)  
%+ \frac{\imgi S B_z}{\sigma}\left(\dvx^{\prime\prime} -
%\frac{\sbar^\prime}{\sbar}\dvx^\prime\right) + \frac{\imgi
%  S}{\sigma}\frac{\mu_0\rho}{B_z} \left(\overline{D}_2 -
%    \frac{\sbar^\prime}{\sbar}\overline{D}_1 \right)U 
%+\frac{\mu_0\rho}{B_z} \left(\overline{D}_2 -
%    \frac{\sbar^\prime}{\sbar}\overline{D}_1 - \imgi\sbar D_0 \right)V. \label{final_vy}
%
%  0 =& \frac{f^2}{\beta\rho}\left(\dvy^{\prime\prime} -
%  \frac{\bar{\sigma}^\prime}{\bar{\sigma}}\dvy^\prime\right) 
%  + \sbar\sigma D_0\dvy  + \frac{\imgi 
%    S}{\sigma}\frac{f^2}{\beta\rho}\left(\dvx^{\prime\prime} -
%    \frac{\sbar^\prime}{\sbar}\dvx^\prime\right) -
%    \frac{\imgi\sbar\kappa^2}{2}D_0\dvx \notag\\
%    & + \left(\overline{D}_2 -
%    \frac{\sbar^\prime}{\sbar}\overline{D}_1\right)\left[\imgi\left(\sigma
%      - \frac{2S}{\sigma}\right)\dvy + \left(\frac{\kappa^2}{2} -
%      S\right)\dvx - \frac{Sfk_x}{\sigma}\w\right]. \label{final_vy}
\end{align}

Eq. \ref{final_vx} and \ref{final_vy} constitutes the first two
linearized equations to be solved.

\subsection{Tilted field with uniform resistivity}
Here we allow $\epsilon\neq 0$ but take $\eta$ to be constant. We
first obtain expressions for $\dbx$ and $\dby$. Differentiating
the $x$ momentum equation and replacing the resulting $\dbz^\prime$
using the divergence-free condition and $\dby^\prime$ using the $y$
momentum equation, we obtain an expression for $\dbx^{\prime\prime}$
which can be inserted into the $x$ induction equation. This gives
\begin{align}
  \imgi\sigma\dbx = B_z\dvx^\prime +
  \frac{\eta\mu_0\rho}{B_z}\left( D_1 U + \imgi\epsilon k_x D_0
  V\right).\label{bx_tilted}  
\end{align}
We can insert this into the $y$ induction equation to obtain
\begin{align}
  \imgi\sbar\dby = -B_y\Delta + B_z\dvy^\prime + \frac{\imgi
    S}{\sigma}\left[ B_z \dvx^\prime+ \frac{\eta\mu_0\rho}{B_z}\left( D_1 U + \imgi\epsilon k_x D_0
  V\right)\right] + \frac{\eta\mu_0\rho}{B_z}D_1V,\label{by_tilted}
\end{align} 
where we have also used the derivative of the $y$ momentum equation to
eliminate  $\dby^{\prime\prime}$. Recall $\Delta \equiv i k_x \dvx +
\dvz^\prime$, so that
\begin{align}
  \Delta = \imgi k_x \dvx + \frac{i}{\sigma}\left(\w^{\prime\prime} +
  \epsilon V^\prime\right) = -\left[\frac{\imgi\sigma W}{c_s^2} +
    \frac{\imgi\left(\ln{\rho}\right)^\prime}{\sigma}\left(\w^\prime +
    \epsilon V\right)\right], 
\end{align}
where the second equality results from the continuity equation. 

Now consider
\begin{align}
  \dbx^\prime - \imgi k_x \dbz = \frac{\mu_0\rho}{B_z}D_0U +
  \imgi\epsilon k_x \dby = \frac{\sigma}{\sbar}\dbx^\prime +
  \frac{\imgi k_x^2B_z}{\sbar}\dvx,
\end{align}
where the first equality corresponds to the $x$ momentum equation and
the second equality results from replacing $\dbz$ using the $z$
induction equation. We can now use the above expressions for $\dbx$
and $\dby$ (Eq. \ref{bx_tilted}--\ref{by_tilted}) to obtain
\begin{align}
0 = &v_A^2\left[k_x^2\left(1+\epsilon^2\right)\dvx - \frac{\epsilon k_x
  S}{\sigma}\dvx^\prime - \dvx^{\prime\prime}\right] + \imgi\epsilon
k_x v_A^2 \dvy^\prime + \frac{\epsilon^2 k_x
  v_A^2}{\sigma}\left(\w^{\prime\prime} + \epsilon V^\prime\right) 
\notag\\ &-\left[\eta\left(D_2 + \frac{\epsilon k_x S}{\sigma}D_1\right) -
  \imgi\sbar D_0 \right]U %\notag\\
- \frac{\imgi\epsilon^2k_x^2S}{\sigma}\eta D_0V.
\end{align}
%We can now use the above expression for $\dbx,\dby$ to eliminate
%$\dbx^\prime$ and $\$ in the $x$ momentum equation  
%\begin{align}
%  \dbx^\prime - \imgi k_x\dbz = \frac{}{}
%\end{align}

Similarly, we differentiate Eq. \ref{by_tilted} and use the $y$
momentum equation to eliminate $\dby^\prime$ to obtain
\begin{align}
0 = &v_A^2\dvy^{\prime\prime} + \frac{\imgi S}{\sigma}v_A^2
\dvx^{\prime\prime} +\frac{\imgi S}{\sigma}\eta D_2U +
\left\{\eta\left(D_2 - \frac{\epsilon k_x S}{\sigma} D_1\right) +
  \imgi\left[\frac{\epsilon^2v_A^2\left(\ln{\rho}\right)^{\prime\prime}}{\sigma}
    - \sbar\right]D_0\right\}V \notag\\ &+ \imgi\epsilon
  v_A^2\left\{\frac{\sigma}{c_s^2}\left[W^\prime -
    \left(\ln{c_s^2}\right)^\prime W\right] +
    \frac{1}{\sigma}\left[\left(\ln{\rho}\right)^\prime(\w^{\prime\prime}
      + \epsilon V^\prime)
      +\left(\ln{\rho}\right)^{\prime\prime}\w^{\prime} \right]\right\}.    
\end{align}

%\subsection{Non-dimensionalization}
%In practice we solve the linearized equations in non-dimensional form,
%by defining
%We further write 
%\begin{align}
%  &z=\hat{z}H,\quad k_x =  \hat{k}_x/H,\quad \sigma = \hat{\sigma}\Omega,
%  \quad \delta\bm{v} = \csmid 
%  \delta\hat{\bm{v}}, \\ 
%  &\delta\bm{B} = B\delta\hat{\bm{B}},\quad
%  \delta\rho = \rho\hat{W}/\hat{c}_s^2,\quad \delta\Phi =
%  \csmid^2\delta\hat{\Phi}, 
%\end{align}
%where we have introduced the non-dimensional enthalpy perturbation
%$\hat{W}$ and the sound-speed $\hat{c}_s=\csmid^{-1}\sqrt{dP/d\rho}$ can be 
%obtained from the equation of state. The background frequencies are
%written $S=\hat{S}\Omega$ and $\Omega_z=\hat{\Omega}_z\Omega$ and the 
%resistivity as $\eta = \hat{\eta}H^2\Omega$. 

%We will now drop the $\hat{\phantom{a}}$ notation. Henceforth it is
%understood that all variables have been appropriately normalized
%(i.e. $\Omega\to 1$ and $H\to 1$), unless otherwise stated.

%tilted field
% calculate b field vertical derivatives carefully
% try no velocity pert at bc
% try anti-aligned field 

%\bibliographystyle{apj}
%\bibliography{ref}

\end{document}